\documentclass[a4paper,fleqn,usenatbib]{mnras}

\usepackage[T1]{fontenc}
\usepackage{ae,aecompl}

\DeclareRobustCommand{\VAN}[3]{#2}
\let\VANthebibliography\thebibliography
\def\thebibliography{\DeclareRobustCommand{\VAN}[3]{##3}\VANthebibliography}

\usepackage{graphicx}	
\usepackage{amsmath}	
\usepackage{amssymb}	

\title[The complex variability of blazars]{The complex variability of blazars: Time-scales and periodicity analysis in S4~0954+65}

\author[C. M. Raiteri et al.] 
{C.~M.~Raiteri              $^{ 1}$\thanks{E-mail:claudia.raiteri@inaf.it} ,
M.~Villata                  $^{ 1}$,
V.~M.~Larionov              $^{ 2,3}$,
S.~G.~Jorstad               $^{ 4,2}$,
A.~P.~Marscher              $^{ 4}$,
Z.~R.~Weaver                $^{ 4}$,
\newauthor
J.~A.~Acosta-Pulido         $^{ 5,6}$,
I.~Agudo                    $^{ 7}$,
T.~Andreeva                 $^{ 8}$,
A.~Arkharov                 $^{ 3}$,
R.~Bachev                   $^{ 9}$,
E.~Ben\'itez                $^{10}$,
\newauthor
M.~Berton                   $^{11,12}$,
I.~Bj\"orklund              $^{12,13}$,
G.~A.~Borman                $^{14}$,
V.~Bozhilov                 $^{15}$,
M.~I.~Carnerero             $^{ 1}$,
\newauthor
D.~Carosati                 $^{16,17}$,
C.~Casadio                  $^{18,19,20}$,
W.~P.~Chen                  $^{21}$,
G.~Damljanovic              $^{22}$,
F.~D'Ammando                $^{23}$,
\newauthor
J.~Escudero                 $^{ 7}$,
A.~Fuentes                  $^{ 7}$,
M.~Giroletti                $^{23}$,
T.~S.~Grishina              $^{ 2}$,
A.~C.~Gupta                 $^{24}$,
V.~A.~Hagen-Thorn           $^{ 2}$,
\newauthor
M.~Hart                     $^{ 4}$,
D.~Hiriart                   $^{25}$,
W.-J.~Hou                   $^{21}$,
D.~Ivanov                   $^{ 8}$,
J.-Y.~Kim                   $^{26}$,
G.~N.~Kimeridze             $^{27}$,
\newauthor
C.~Konstantopoulou          $^{28,29}$,
E.~N.~Kopatskaya            $^{ 2}$,
O.~M.~Kurtanidze            $^{27,30,31}$,
S.~O.~Kurtanidze            $^{27,30,32}$,
\newauthor
A.~L\"ahteenm\"aki          $^{12,13}$,
E.~G.~Larionova             $^{ 2}$,
L.~V.~Larionova             $^{ 2}$,
N.~Marchili                 $^{23}$,
G.~Markovic                 $^{33}$,
\newauthor
M.~Minev                    $^{15}$,
D.~A.~Morozova              $^{ 2}$,
I.~Myserlis                 $^{34,20}$,
M.~Nakamura                 $^{35}$,
A.~A.~Nikiforova            $^{ 3,2}$,
\newauthor
M.~G.~Nikolashvili          $^{27,30}$,
J.~Otero-Santos             $^{ 6,5}$,
E.~Ovcharov                 $^{15}$,
T.~Pursimo                  $^{28}$,
I.~Rahimov                  $^{ 8}$,
\newauthor
S.~Righini                  $^{23}$,
T.~Sakamoto                 $^{35}$,
S.~S.~Savchenko             $^{ 2,36,3}$,
E.~H.~Semkov                $^{ 9}$,
D.~Shakhovskoy              $^{14}$,
\newauthor
L.~A.~Sigua                 $^{27}$,
M.~Stojanovic               $^{22}$,
A.~Strigachev               $^{ 9}$,
C.~Thum                     $^{34}$,
M.~Tornikoski               $^{12}$,
E.~Traianou                 $^{20}$,
\newauthor
Y.~V.~Troitskaya            $^{ 2}$,
I.~S.~Troitskiy             $^{ 2}$,
A.~Tsai                     $^{21}$,
A.~Valcheva                 $^{15}$,
A.~A.~Vasilyev              $^{ 2}$,
O.~Vince                    $^{22}$,
\newauthor
and E.~Zaharieva             $^{15}$\\
{\it Affiliations are listed at the end of the paper}
}

\date{Accepted XXX. Received YYY; in original form ZZZ}

\pubyear{2020}

\begin{document}
\label{firstpage}
\pagerange{\pageref{firstpage}--\pageref{lastpage}}
\maketitle

\begin{abstract}
Among active galactic nuclei, blazars show extreme variability properties. 
We here investigate the case of the BL Lac object S4~0954+65 with data acquired in 2019--2020 by the {\it Transiting Exoplanet Survey Satellite} ({\it TESS}) and by the Whole Earth Blazar Telescope (WEBT) Collaboration. The 2-min cadence optical light curves provided by {\it TESS} during three observing sectors of nearly one month each, allow us to study the fast variability in great detail. We identify several characteristic short-term time-scales, ranging from a few hours to a few days. However, these are not persistent, as they differ in the various {\it TESS} sectors. The long-term photometric and polarimetric optical and radio monitoring undertaken by the WEBT brings significant additional information, revealing that i) in the optical, long-term flux changes are almost achromatic, while the short-term ones are strongly chromatic; ii) the radio flux variations at 37 GHz follow those in the optical with a delay of about three weeks; iii) the range of variation of the polarization degree and angle is much larger in the optical than in the radio band, but the mean polarization angles are similar; 
iv) the optical long-term variability is characterized by a quasi-periodicity of about one month.
We explain the source behaviour in terms of a rotating inhomogeneous helical jet, whose pitch angle can change in time. 

\end{abstract}




\section{Introduction}
Blazars are active galactic nuclei that show extreme variability properties. 
They include flat-spectrum radio quasars, generally showing broad emission lines in their spectra, and (almost) featureless BL Lac objects. 
Blazar emission mostly comes from a relativistic plasma jet that is oriented closely to the line of sight. 
As a consequence, the flux is Doppler beamed and boosted, and time-scales are shortened. 
Blazar variability is unpredictable. Some objects show almost continuous activity, while others undergo extreme outburst events after periods of almost constant emission.
Variability time-scales range from years down to hours, likely implying that different mechanisms are at work, both of intrinsic (i.e. energetic) and extrinsic (i.e. geometric) nature.

The {\it Transiting Exoplanet Survey Satellite} \citep[{\it TESS};][]{ricker2015} that was launched in 2018, though dedicated to the discovery of exoplanets, gives us the possibility to study the short-term blazar variability with extremely dense sampling. Therefore, we proposed a number of bright blazars for {\it TESS} observations in cycles 2, 3, and 4.
Some of these {\it TESS} blazar observations were supported by ground-based monitoring by the Whole Earth Blazar Telescope\footnote{http://www.oato.inaf.it/blazars/webt/} (WEBT) Collaboration \citep[e.g.][]{villata2002,villata2004,villata2006,villata2008,raiteri2013,raiteri2017_nature,carnerero2015,carnerero2017,larionov2016,larionov2020}, which in particular added colour information.

The 2-min cadence {\it TESS} observations of BL Lacertae were analysed by \citet{weaver2020}, together with data from the WEBT, and the {\it Neil Gehrels Swift}, the {\it Nuclear Spectroscopic Telescope Array} ({\it Nustar}), and the {\it Fermi} satellites. The authors found that the shortest variability time-scale in the {\it TESS} light curve is about 0.5 hr, while the most common time-scale is around 13 hr. This is similar to the minimum time-scale identified in the X-rays, 14.5 hr. 
They found a statistically significant correlation between the {\it TESS} and X-ray light curves, with X-ray variations lagging the optical ones up to $\sim 0.4$ d, and explained the source behavior as due to a shock propagating in turbulent plasma.

In \cite{raiteri2021} we analysed the short-term optical variability of the BL Lac object S5~0716+714 with both {\it TESS} and WEBT data. We found that the flux variations of this source have a dual nature. Variability on time-scales shorter than about 5 h shows a clear bluer-when-brighter behaviour and is likely due to energetic processes in jet regions of milliparsec size.
Flux changes on time-scales longer than half a day are instead almost achromatic. We proposed a geometrical explanation for them. Since there are various variability time-scales, we suggested that the jet is made up of filaments twisting around the jet axis. Each filament has its own changing orientation with respect to the line of sight and thus its emission is affected by a variable Doppler factor.

We here apply the same methods used in that paper to the {\it TESS} and WEBT observations of another BL Lac object, S4~0954+65. 
The redshift of this source is still debated.
A tentative estimate $z=0.368$ was given by \citet{lawrence1986} based on [\ion{O}{III}] $\lambda \lambda 4959, 5007$ and galactic absorption lines. The value was essentially confirmed by \citet{stickel1993}, who determined $z=0.367$ from the identification of \ion{Ca}{II} $\lambda \lambda 3933,3968$ and extremely weak [\ion{O}{II}] $\lambda 3727$.
These redshift values have been questioned by \citet{landoni2015}, who claimed a lower limit of 0.45.
However, recently \citet{becerra2020} gave a more precise estimate, $z=0.3694\pm0.0011$, using the \ion{Mg}{II} $\lambda 2800$ emission line visible during a faint state of the source\footnote{They also identified the [\ion{O}{III}], and possibly the [\ion{O}{II}] emission lines.}. 

S4~0954+65 is well-known for its fast flux changes. Intranight/intraday variability in the optical/radio band has been detected many times, especially when the source was in flaring state \citep{wagner1990,wagner1993,heidt1996,raiteri1999,marchili2012,morozova2014,bachev2015}.
In some of these studies, geometrical effects were invoked to explain at least part of the variability.
During the 2008 radio outburst, the intraday variability properties were seen to change, with the appearance of multiple time-scales \citep{marchili2012}.
The source was first detected at very high energies ($\ge 100 \, \rm GeV$) during an exceptionally bright state in February 2015 \citep{ahnen2018}. Both the 2011 and 2015 outbursts were accompanied by wide rotations of the optical polarization angle and they were associated with the emission of new jet components observed in high-resolution radio images \citep{morozova2014,ahnen2018}.
The swing of the polarization angle observed in 2011, from 0\degr\ to 330\degr\ and then back to 0\degr, was interpreted by \citet{lyutikov2017} in terms of an oscillating circular motion.

Differently from the work done on S5~0716+714 by \citet{raiteri2021}, who analysed about one month of {\it TESS} and WEBT data, in this paper we explore {\it TESS} data coming from three observing sectors, lasting about one month each, and WEBT monitoring data extending to the whole 2019--2020 optical observing season. Moreover, WEBT data include now also radio photometric and radio and optical polarimetric observations. In this way we can investigate both short and long variability time-scales, from a few hours to several weeks, and put them in context, considering the source spectral and polarization variability.

The paper is structured as follows. In Section~\ref{sec_tess} we describe the {\it TESS} observations of S4~0954+65 during cycle 2.
The data provided by the WEBT are presented in Section~\ref{sec_webt} (optical and near-infrared), and Section~\ref{sec_radio} (radio). The behaviour of colour indices is discussed in Section~\ref{sec_col} and those of the polarization degree and angle in Section~\ref{sec_pola}. A detailed analysis of the short-term variability in the {\it TESS} light curves is performed in Section~\ref{sec_stv}, using the autocorrelation function (and structure function) and periodogram. The long-term variability in one year of WEBT and {\it TESS} observations is studied in Section~\ref{sec_ltv}. A rotating helical jet model is applied to the optical observations in Section~\ref{sec_mod}. The results are summarized in Section~\ref{sec_fine}, where the conclusions are finally drawn.

\section{Observations by TESS}
\label{sec_tess}
Blazar S4~0954+65 was observed by the {\it TESS} satellite during its cycle 2, targeting the northern ecliptic sky. 
The source was monitored in Sector 14 (2019 July 18.84 -- August 14.70), and then in Sector 20 (2019 December 25.01 -- 2020 January 20.33) and 21 (2020 January 21.93 -- 2020 February 18.28). 

\begin{figure}
	\includegraphics[width=\columnwidth]{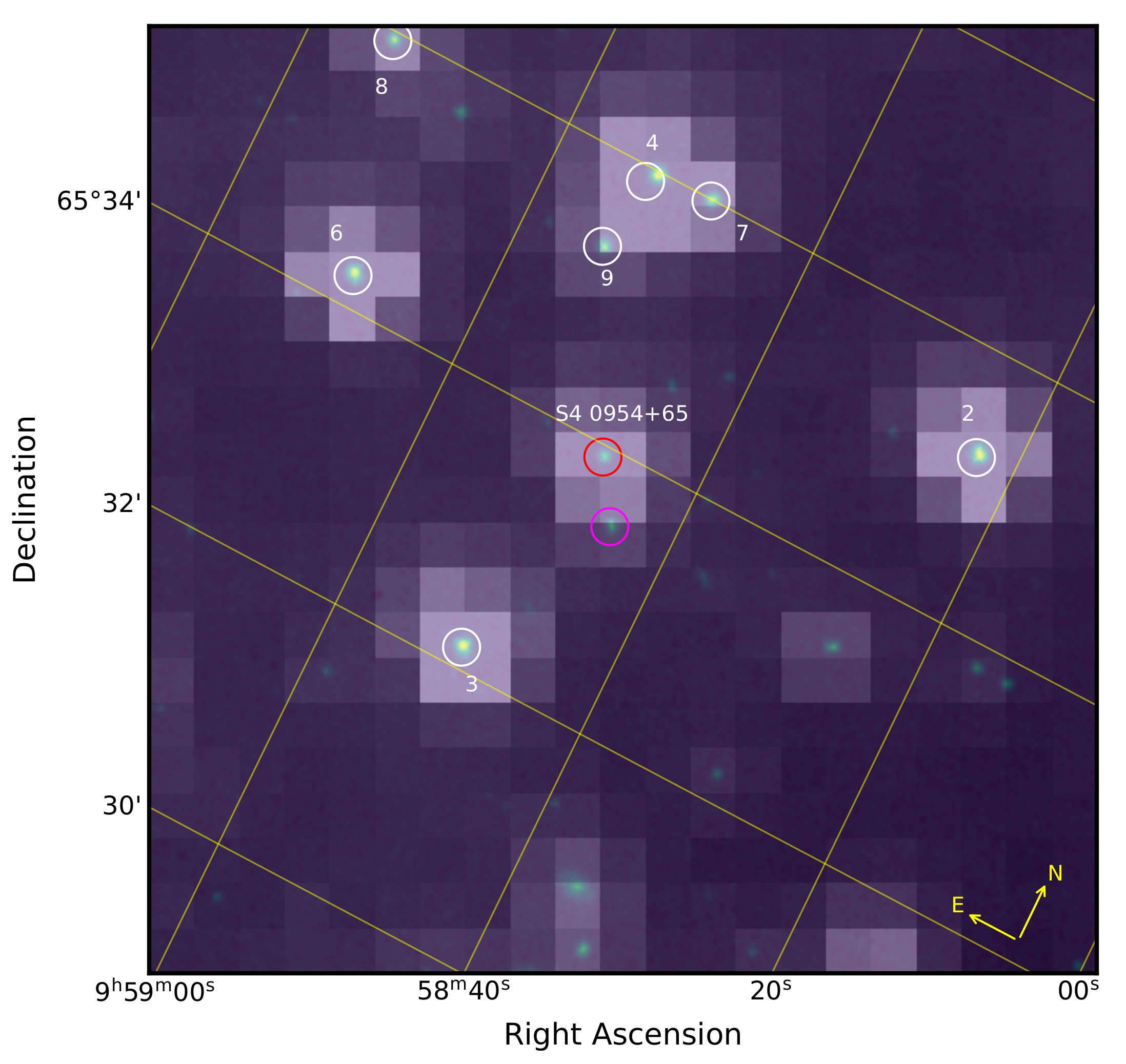}
    \caption{A $21 \times 21$ pixel cutout of a {\it TESS} Full Frame Image (FFI) centered on S4~0954+65 (red circle) from 2019 August 09 16:15:36 UT (large pixels). A Digitized Sky Survey 2 (red filter) image of the same field is shown in the background. Close comparison stars are circled in white and labelled according to \citet{raiteri1999}, while the nearby faint source is circled in magenta. Differences in the {\it TESS} locations (circles) and DSS PSFs can be reconciled by the proper motion of the sources.}
    \label{fc}
\end{figure}

Given the large pixel size of the {\it TESS} detector, crowded fields may complicate the analysis of blazar light curves. Fig.~\ref{fc} shows a cutout of a {\it TESS} Full Frame Image (FFI) taken during Sector~14 and centred on S4~0954+65. All main comparison stars from \citet{raiteri1999} are sufficiently far from the source and do not have PSFs that overlap. 
A single faint star to the southwest does overlap with the blazar in the algorithm-determined 2-min cadence aperture of each observing sector. However, taking the PSF of the source into account, this aperture captures $\sim 2/3$ of the total estimated flux from S4~0954+65. Given that the blazar is relatively isolated from nearby stars and that the majority of light in the automatically-determined aperture comes from the source, we conclude that there are no systematic issues due to flux contamination to correct for.

\begin{figure}
	\includegraphics[width=\columnwidth]{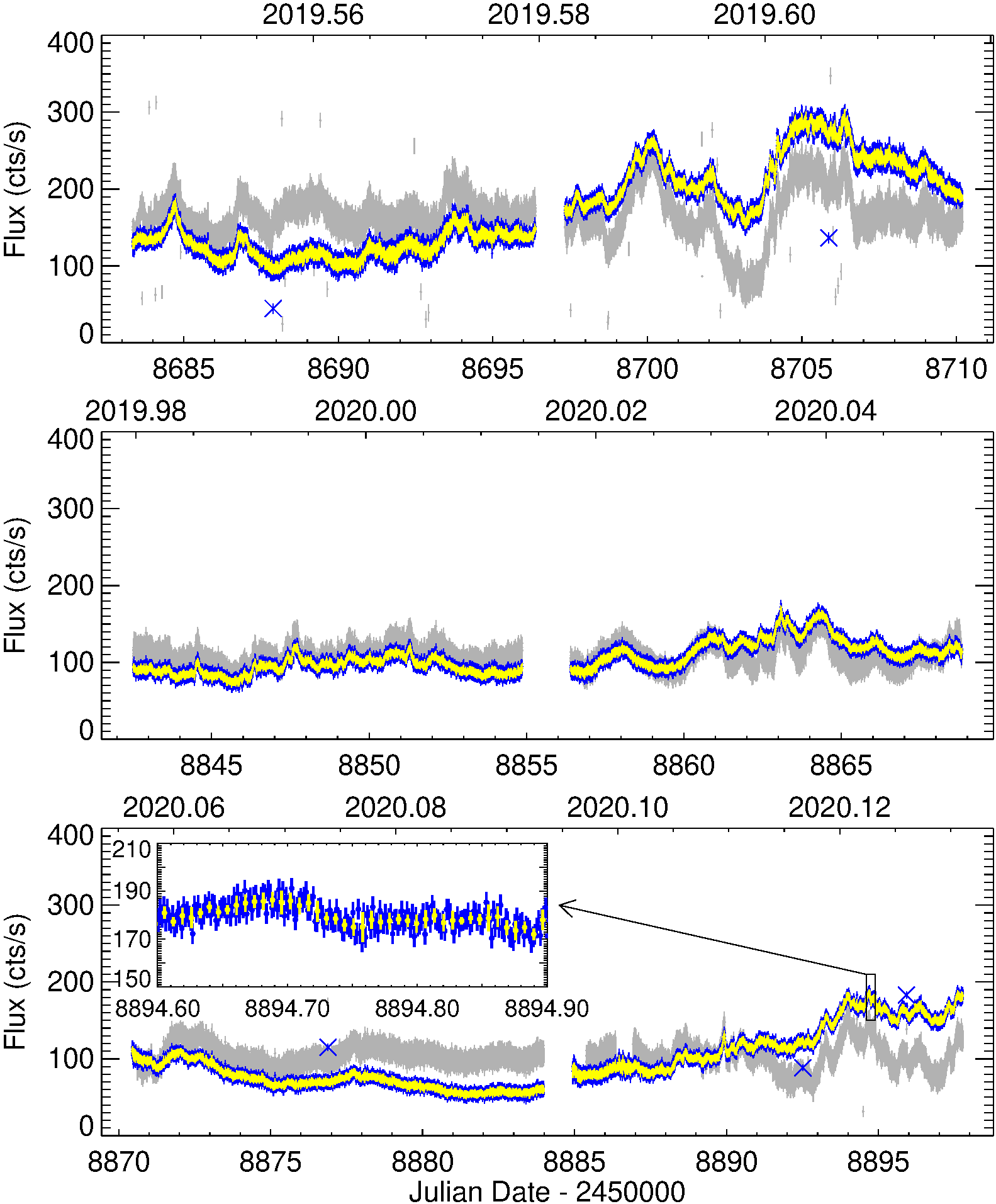}
    \caption{{\it TESS} light curves of S4~0954+65 during Sector 14 (top), 20 (middle) and 21 (bottom). Fluxes are given as electrons per second. Blue dots and bars represent the 2-min cadence simple-aperture photometry fluxes (SAP\textunderscore FLUX) and errors; blue crosses indicate the outliers that have been removed before binning; yellow points and bars show the 10-min binned fluxes and their standard deviations. Grey dots and bars represent conditioned fluxes (PDCSAP\textunderscore FLUX) and errors. In the inset panel a zoom into the Sector 21 light curve shows in more details the effect of the 10-min binning.}
    \label{tess}
\end{figure}

The 2-min cadence light curves are shown in Fig.~\ref{tess}. They were downloaded from the Mikulski Archive for Space Telescopes\footnote{https://mast.stsci.edu/portal/Mashup/Clients/Mast/\\Portal.html}.
There is a remarkable difference in trend between the simple-aperture photometry fluxes (SAP\textunderscore FLUX) and the pre-search data conditioned simple aperture photometry fluxes (PDCSAP\textunderscore FLUX). 
With respect to the former, the latter were further processed to remove possible instrumental effects that are known to affect light curves of stars that are exoplanets host candidates. This translates into removing long-term trends. However, this process may erase true source variability in objects like blazars.
Moreover, the PDCSAP\textunderscore FLUX data points show much more scatter and outliers, especially during Sector 14.
We will show in Section~\ref{sec_webt} that ground-based optical data confirm the reliability of the SAP\textunderscore FLUX light curves.

Because of the source relative faintness, the {\it TESS} light curves are somewhat noisy.
Therefore, we first removed the most evident outliers by discarding those data points that are out of 4 standard deviations in the 0.1 d binned light curve. Then we binned the SAP fluxes in time intervals of 10 min and assumed the standard deviation in each bin as the data uncertainty (see Fig.~\ref{tess}).
In the rest of the paper we will use these 10-min binned SAP fluxes, including 11097 data points.

\section{Optical and near-infrared observations by the WEBT}
\label{sec_webt}
To increase the scientific return of the {\it TESS} observations, the WEBT Collaboration organized a multiwavelength monitoring campaign. 
In this section, we present the optical and near-infrared data acquired by the WEBT over a whole year including the {\it TESS} pointings, from 2019 July 18 to 2020 July 18 ($\rm JD=2458683$--2459049). 
Table~\ref{obs} lists the participating observatories.

In the optical band, 1764 data in the $BVRIz$ filters were taken with 20 telescopes around the world, with diameters from 35 to 260 cm. Nearly 64\% of the optical data are in the Cousins' $R$ filter, which will be considered our reference band.
Data were reduced according to standard prescriptions.
The source magnitude was calibrated with respect to the photometric sequence by \citet{raiteri1999} in $BVRI$ and using PanSTARRS\footnote{https://panstarrs.stsci.edu/} PS1 magnitudes for the few data acquired with the Nordic Optical Telescope (NOT) in the Sloan Digital Sky Survey\footnote{https://www.sdss.org/} (SDSS) $z$-band. 
A few near-infrared observations were done with the 2~m Liverpool telescope in $H$ band. 
Calibration of these data was obtained using Two Micron All Sky Survey\footnote{https://irsa.ipac.caltech.edu/Missions/2mass.html} (2MASS) sources in the field.

\begin{figure}
	\includegraphics[width=\columnwidth]{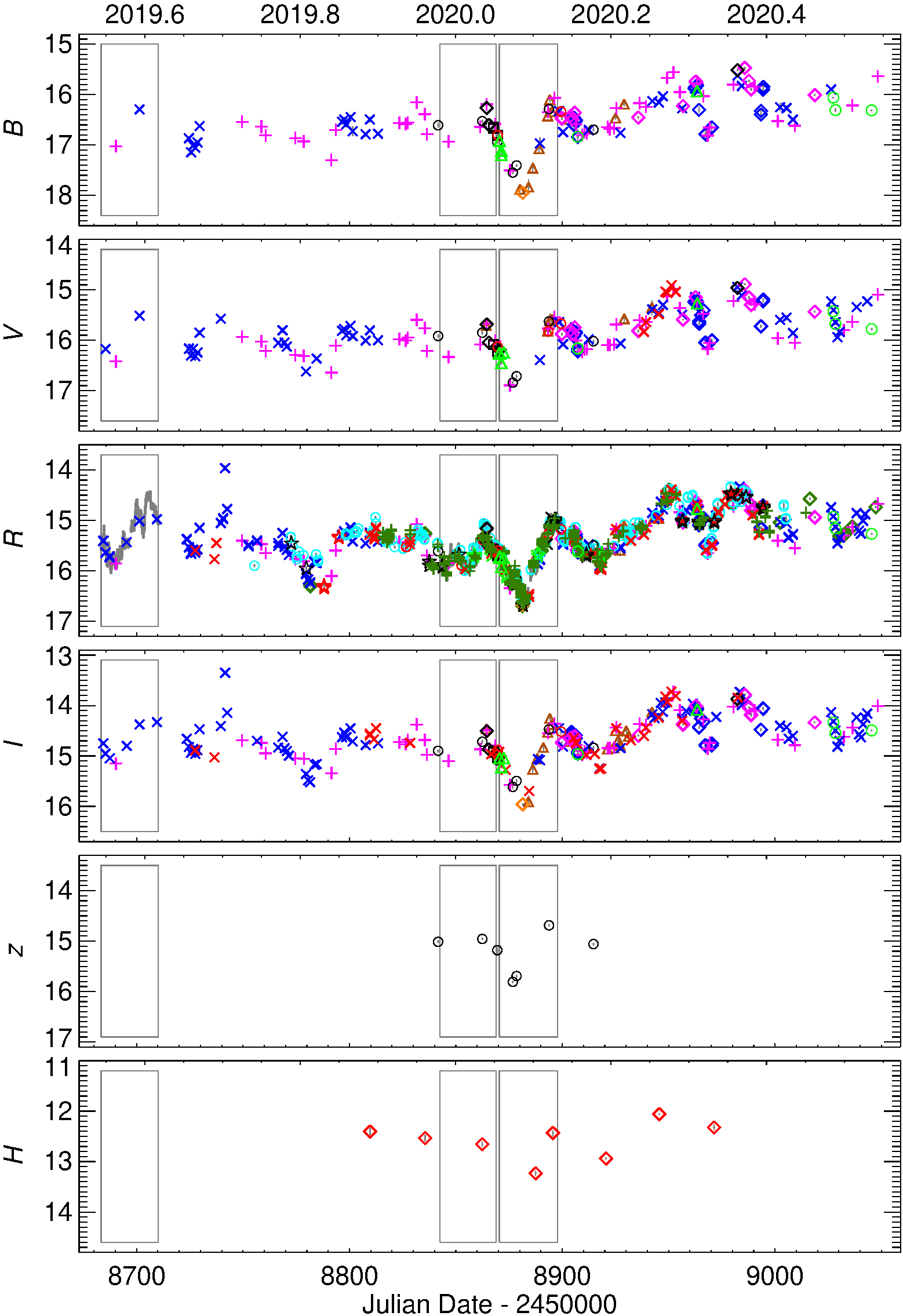}
    \caption{Multiband optical light curves of S4~0954+65 by the WEBT Collaboration. Different symbols and colours distinguish different observatories and telescopes, as listed in Table~\ref{obs}. The grey boxes mark the periods of {\it TESS} observations and the grey dots in the $R$-band panel represent the {\it TESS} SAP fluxes converted into $R$-band magnitudes as explained in the text.
}
    \label{ottico}
\end{figure}

In the $BVRI$ bands, the different datasets were combined and the resulting multiband light curves were visually inspected and processed to obtain homogeneous and reliable curves for further analysis.
We eliminated clear outliers and reduced data noise by binning data close in time from the same telescope, when appropriate. 
In only three cases we had to apply shifts to align the datasets with the trend traced by all the others.
The final optical light curves are shown in Fig.~\ref{ottico}. 
They reveal continuous activity, with variability amplitudes ($\rm max -min $) of 2.48, 2.00, 2.82, 2.61  mag in $BVRI$ bands, respectively.
Although much less sampled, the magnitude range spanned in the $z$ and $H$ bands are 1.12 and 1.18, respectively.

We note that the larger values in the $R$ and $I$ bands are due to a remarkable variability episode occurred on $\rm JD=2458740$--42, when the source brightened by 1.0 mag in 24 h in the $R$ band, and then dimmed by 0.8 mag in 23 h. We carefully checked the reliability of this episode, which is also visible in the $I$ band, and found no reasons to distrust it.
Though rather extreme, this event is not unique, as e.g.\ a brightening of about 0.9 mag in 24 h was observed by \citet{raiteri2009} in 2007, during a WEBT campaign targeting BL Lacertae.
Without this episode, the variability amplitudes of S4~0954+65 in $R$ and $I$ bands are 2.46 and 2.24 mag, respectively.
 
\begin{figure}
	\includegraphics[width=\columnwidth]{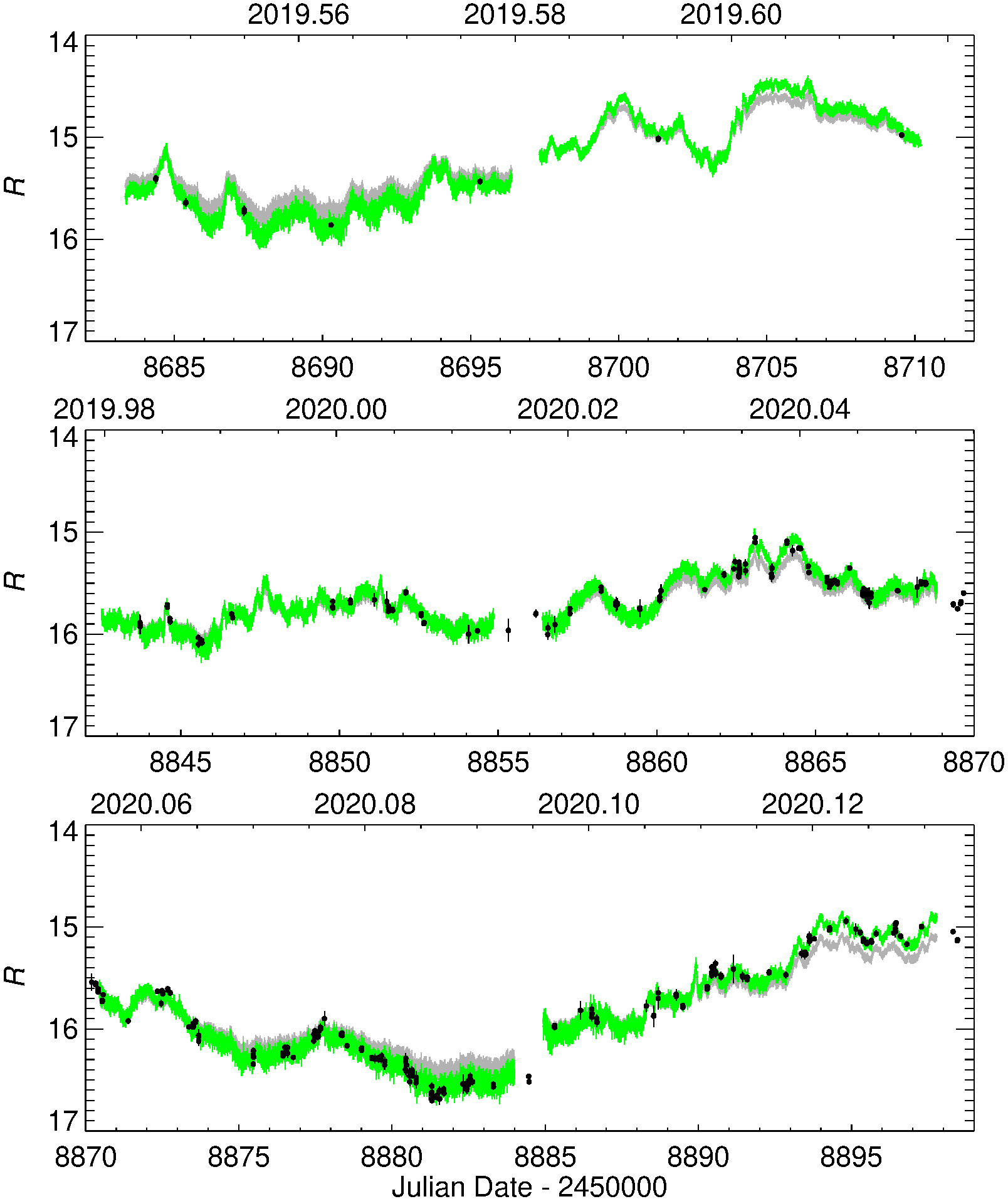}
    \caption{The $R$-band light curve in {\it TESS} Sector 14 (top), Sector 20 (middle) and Sector 21 (bottom). Black dots represent WEBT data. Grey dots show {\it TESS} SAP fluxes converted into magnitudes according to the simple formula $\rm mag_{\it TESS}=-2.5 \, \log $ (SAP\textunderscore FLUX)$\rm + mag_0$, with $\rm mag_0=20.75$. Because of the {\it TESS} spectral response, this is not able to reproduce the $R$-band variability amplitude. Green dots show the {\it TESS} fluxes converted into $R$-band magnitudes according to $\rm mag_{\it TESS}=-2.5 \, \log $ (SAP\textunderscore FLUX)$^\zeta \rm + mag_0$, with best-fit values $\zeta=1.27$ and $\rm mag_0=22.08$ in Sectors 20+21, and $\rm mag_0=22.28$ in Sector 14.
}
    \label{stretch}
\end{figure}

Fig.~\ref{ottico} also shows the periods of {\it TESS} observations and the {\it TESS} fluxes converted into $R$-band magnitudes. 
These {\it TESS} magnitudes adapted to the reference $R$ band will be used in the following to perform time series analysis on the composite light curve, including both {\it TESS} and WEBT data.
The transformation of the {\it TESS} fluxes into magnitudes was done by taking into account that the {\it TESS} detector bandpass\footnote{https://heasarc.gsfc.nasa.gov/docs/tess/the-tess-space-telescope.html}, 
which is very wide, spanning from 600 to 1000 nm, is centered on the Cousins' $I$ band. 
As we saw above, the variability amplitude in S4~0954+65 is likely to increase from the $I$ to the $R$ band, so 
a simple flux-to-magnitude formula $\rm mag_{\it TESS}=-2.5 \, \log $ (SAP\textunderscore FLUX)$\rm + mag_0$ cannot reproduce the whole magnitude range observed in the $R$ band (see Fig.~\ref{stretch}, where $\rm mag_0$ is set to 20.75 to obtain a good match). 
Therefore, we first ``stretched" the {\it TESS} fluxes and then converted them into magnitudes, according to: $\rm mag_{\it TESS}=-2.5 \, \log $ (SAP\textunderscore FLUX)$^\zeta \rm + mag_0$ \citep[see also][]{raiteri2021}.
To find the best-fit values for $\zeta$ and $\rm mag_0$, we built a grid of $100 \times 100$ {\it TESS} light curves, changing $\zeta$ and $\rm mag_0$ in steps of 0.01, and found the case which minimized the root mean square of the mag differences between the closest WEBT and {\it TESS} data points.
The {\it TESS} observations in Sector 14 occurred near the source solar conjunction and only a few WEBT data are available; those in Sectors 20 and 21 were supported by the WEBT core campaign, during which an exceptional sampling was reached in the $R$ band. 
For the best-sampled period of Sector 20+21, we got $\zeta=1.27$ and $\rm mag_0=22.08$. As for Sector 14, we fixed $\zeta=1.27$ and derived $\rm mag_0=22.28$. The reason why there is an offset of 0.2 mag between Sector 14 and Sector 20+21 is due to {\it TESS} calibration problems, particularly to the fact that the source was imaged in Camera 4 during Sector 14, but in Camera 2 in Sector 20+21.
The result of our procedure is shown in Fig.~\ref{stretch} and highlights the excellent match between {\it TESS} and WEBT data obtained.
Incidentally, the WEBT data confirm that in the case of S4~0954+65 the SAP fluxes are the right quantities to consider \citep[see also][]{raiteri2021} instead of the PDCSAP fluxes.

\begin{table*}
	\centering
	\caption{Observatories contributing data to the WEBT campaign on S4~0954+65.}
	\label{obs}
	\begin{tabular}{llrccl}
		\hline
		Observatory & Country &  Tel.~size & Band & Symbol & Colour\\
                            &         &  (cm)      &      &        &       \\
		\hline
                \multicolumn{6}{c}{\it Optical}\\
		Abastumani       & Georgia     & 70    & $R$    & {\LARGE $\diamond$} & dark green\\
		Aoyama-Gakuin    & Japan       & 35    & $BVRI$ & {$\triangle$}       & brown\\
		ARIES            & India       & 130   & $BRI$  & {\LARGE $\diamond$} & orange\\
                Belogradchik     & Bulgaria    & 60    & $BVRI$ & {\LARGE $\diamond$} & blue\\
		Crimean$^a$      & Russia      & 70    & $BVRI$ & {\large $\times$}   & blue\\
		Crimean$^b$      & Russia      & 70    & $BVRI$ & {\large $+$}        & magenta\\
		Crimean$^c$      & Russia      & 260   &        & {\LARGE $\diamond$} & green\\
                Haleakala$^d$    & US          & 40    & $R$    & {\LARGE $\star$}    & red\\ 
		Lulin            & Taiwan      & 40    & $R$    & {\LARGE $\circ$}    & cyan\\
                McDonald$^d$     & US          & 40    & $R$    & {\LARGE $\star$}    & green\\
                Perkins          & US          & 180   & $BVRI$ & {\LARGE $\diamond$} & magenta\\
		Roque de los     & Spain       & 260   & $BVRIz$& {\LARGE $\circ$}    & black\\
                Muchachos$^e$    &             &       &        &                     &  \\
		Rozhen           & Bulgaria    & 50/70 & $BVRI$ & {\LARGE $\diamond$} & black\\
		Rozhen           & Bulgaria    & 200   & $BVRI$ & $\square$           & black\\
		San Pedro Martir & Mexico      & 84    & $R$    & {\large $\times$}   & black\\
		St.~Petersburg   & Russia      & 40    & $BVRI$ & {\large $\times$}   & red\\
		Teide$^d$        & Spain       & 40    & $R$    & {\LARGE $\star$}    & black\\
		Teide            & Spain       & 80    & $BVRI$ & {\LARGE $\circ$}    & red\\
		Tijarafe         & Spain       & 40    & $R$    & {\large $+$}        & dark green\\
                Vidojevica$^f$   & Serbia      & 60    & $BVRI$ & {$\triangle$}       & green\\
                Vidojevica$^f$   & Serbia      & 140   & $BVRI$ & {\LARGE $\circ$}    & green\\
                                 &             &       &        &                     &      \\

                \multicolumn{6}{c}{\it Near-infrared}\\
 		Roque de los     & Spain       & 200   & $H$    & {\LARGE $\diamond$} & red \\
                Muchachos$^g$    &             &       &        &                     &  \\
               
                \hline
                Observatory & Country    &  Tel.~size  & Band & Symbol & Colour\\
                            &            &  (m)        & (GHz)&        &       \\
                \hline
                \multicolumn{6}{c}{\it Radio}\\
		Cagliari$^h$    & Italy  & 64          &   6.0      & {\large $+$} & magenta\\
                Cagliari$^h$    & Italy  & 64          &   24       & {$\triangle$} & black\\
                Medicina        & Italy  & 32          &   8.5      & {\large $\times$} & black\\
                Mets\"ahovi     & Finland & 14         &   37       & {\LARGE $\diamond$} & blue \\
		Pico Veleta     & Spain  & 30          &   230      & $\square$   & blue\\
                Pico Veleta     & Spain  & 30          &    86      & {\LARGE $\circ$} & black\\
                Svetloe         & Russia & 32          &   8.5      & {\large $\times$} & blue\\
                Svetloe         & Russia & 32          &   4.8      & {\LARGE $\circ$} & orange\\

		\hline
	\end{tabular}\\
$^a$~ST-7 detector; 
$^b$~AP7p detector; 
$^c$~Only polarization; 
$^d$~Within Las Cumbres Observatory global telescope network;
$^e$~Nordic Optical Telescope (NOT); 
$^f$~Astronomical Station Vidojevica; 
$^g$~Liverpool Telescope; 
$^h$~Sardinia Radio Telescope (SRT)\\
\end{table*}

\section{Optical colours}
\label{sec_col}
Blazars are known also for their spectral variability, which often shows trends with brightness.
In particular, BL Lac objects usually follow a bluer-when-brighter behaviour, especially during fast flares \citep[e.g.][]{villata2004,wu2005,larionov2010,cheng2013,agarwal2015,hagen2015,raiteri2015,raiteri2021}.
In \citet{raiteri2021}, we found that the short-term variability of another BL Lac object, S5~0716+714, presents a double nature. Flux changes on time-scales less than a few hours are strongly chromatic and 
are likely due to energetic processes occurring in small (milliparsec scale) jet zones, while those on time-scales longer than about half a day are quasi-achromatic and are most likely the consequence of changes in the orientation of the jet emitting regions with respect to the line of sight.

To analyse the spectral optical variability of S4~0954+65 in the period considered in this paper, we combine the WEBT $BVRI$ data to build colour indices. The prescription is to use pairs of good data (with uncertainties less than 0.03 mag) acquired from the same telescope within 20 min. 
We obtained 82 $B-R$, 159 $R-I$, and 123 $V-I$ colour indices, with average values of $1.14 \pm 0.08$, $0.68 \pm 0.04$, and $1.20 \pm 0.06$, respectively.
From the $V-I$ index, by transforming magnitudes into de-reddened flux densities\footnote{Adopting the absolute fluxes by \citet{bessell1998} and the values for the Galactic extinction from the NASA/IPAC Extragalactic Database, https://ned.ipac.caltech.edu/} and adopting a power-law $F_\nu \propto \nu^{-\alpha}$ to describe the spectral shape, we obtain an average index $\alpha=1.82 \pm 0.15$, which implies a steep optical spectrum.

\begin{figure}
	\includegraphics[width=\columnwidth]{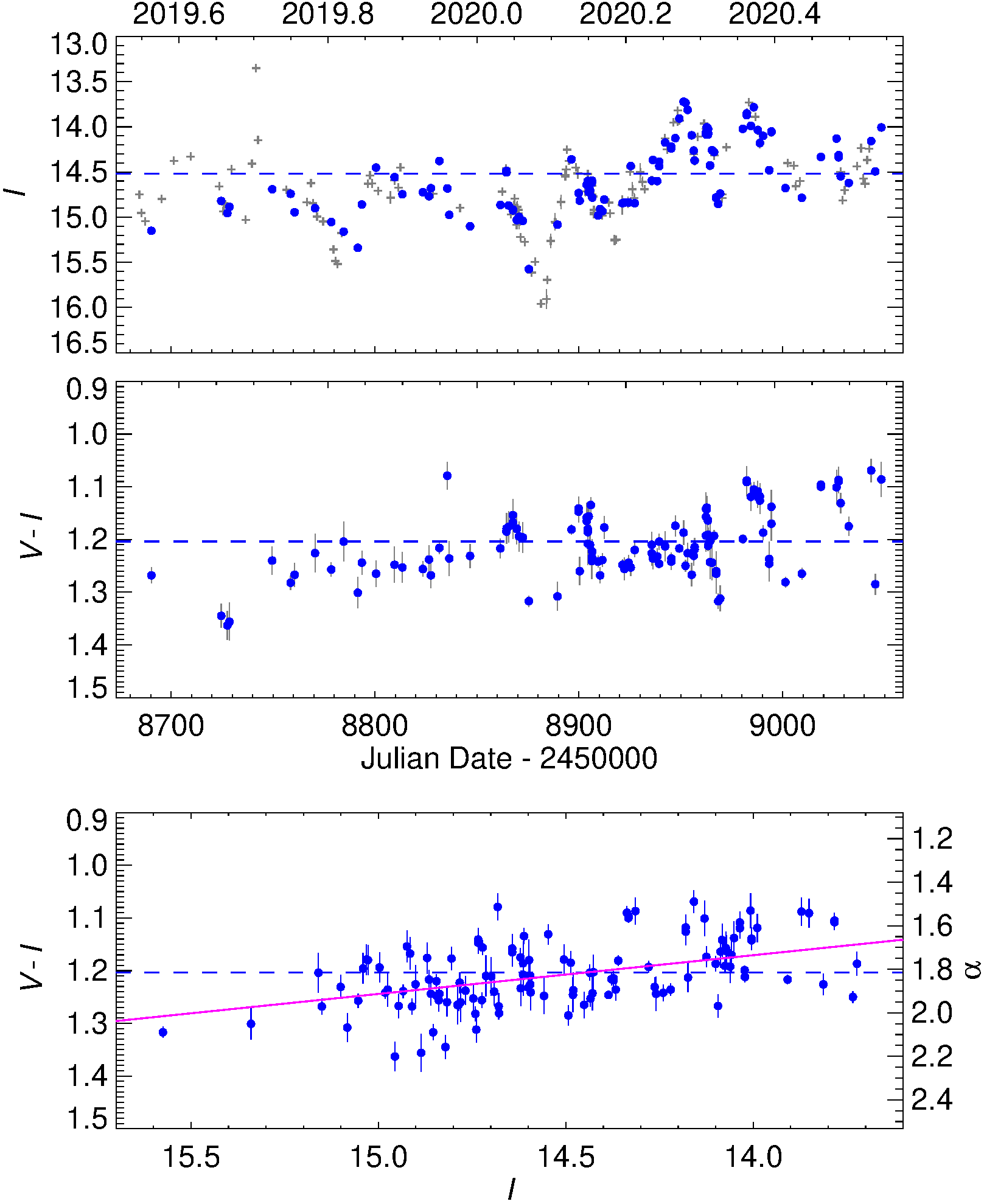}
    \caption{The top panel shows the light curve of S4~0954+65 in the $I$ band already displayed in Fig.~\ref{ottico} (grey plus signs). The blue dots highlight those points used to build the $V-I$ colour indices, which are plotted in the middle panel as a function of time and in the bottom panel as a function of brightness. Horizontal dashed lines indicate average values. The linear fit (magenta line) in the bottom panel suggests a mild bluer-when-brighter trend. The $\alpha$ parameter on the right is the optical spectral index of the $F_\nu \propto \nu^{-\alpha}$ law derived from the $V-I$ colour.
}
    \label{colori}
\end{figure}

Fig.~\ref{colori} compares the behaviour of brightness in the $I$ band and $V-I$ colour as a function of time, and shows the colour (and spectral index) versus brightness. A linear fit indicates a slope of 0.07 and the correlation coefficient is only 0.52. As a consequence, we can say that the long-term spectral variability of S4~0954+65 is only mildly chromatic, with a hint of a bluer-when-brighter behaviour.
These quasi-achromatic variations can be explained by changes of the Doppler factor, if the optical spectral slope is approximately constant. Alternatively, they can be ascribed to a change in the plasma density, or to a variation in the magnetic field strength, again if the optical spectral index were roughly constant, and radiative losses did not increase.
In contrast, the short-term brightness changes display a pronounced chromatism, as most notably visible around $\rm JD-2450000=8905$ and 8965 in Fig.~\ref{colori}. The bad linear fit itself is the consequence of these fast chromatic variations.

\section{Radio observations by the WEBT}
\label{sec_radio}

\begin{figure}
	\includegraphics[width=\columnwidth]{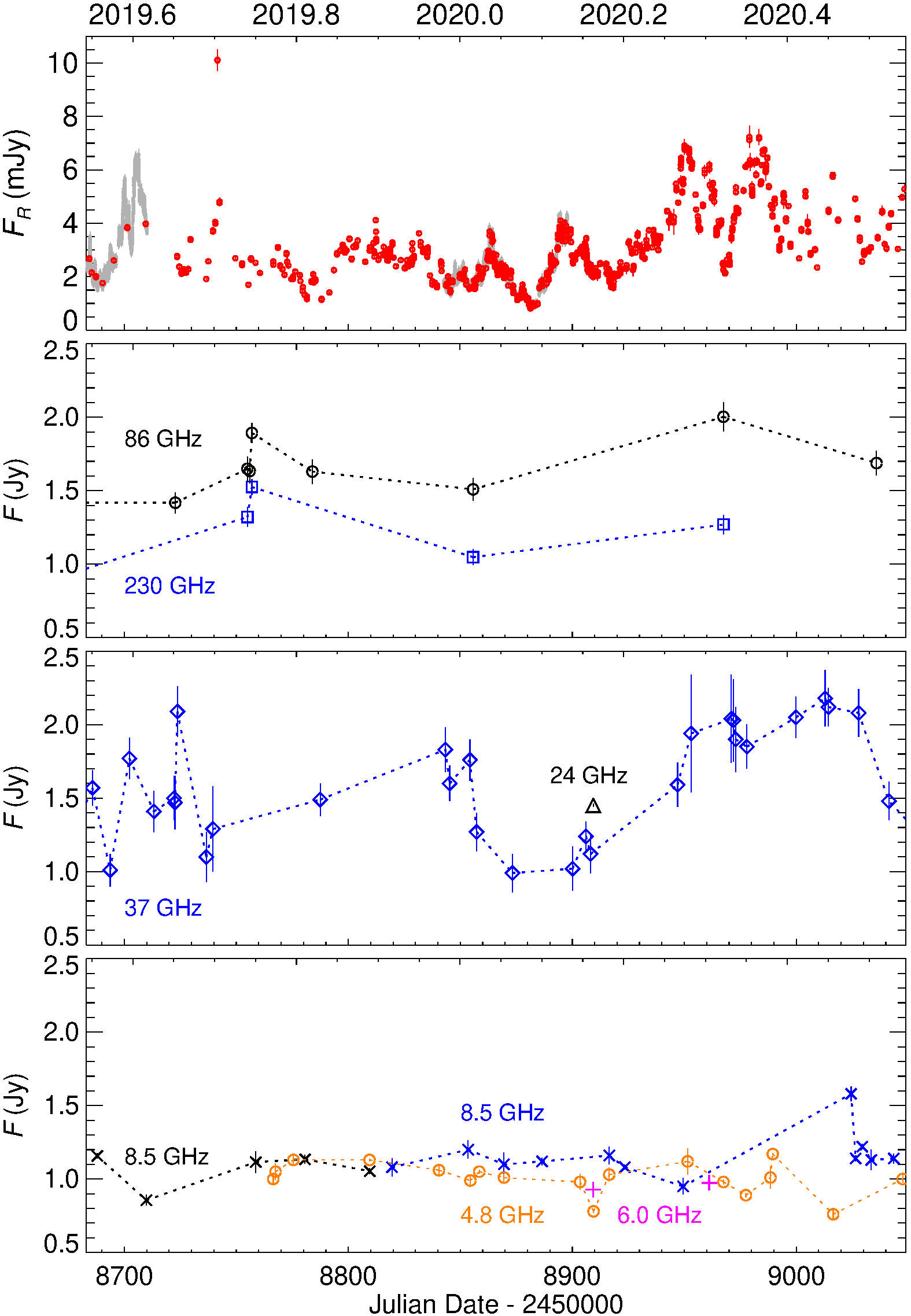}
    \caption{De-reddened WEBT (red dots) and {\it TESS} (grey dots) flux densities in the $R$ band,
compared to radio light curves at various frequencies, decreasing from top to bottom.
Different symbols and colours distinguish different observatories and bands, as listed in Table~\ref{obs}. 
}
    \label{radop}
\end{figure}

Radio data from 4.8 to 37 GHz were obtained with four telescopes (see Table~\ref{obs}) and reduced with standard procedures \citep{terasranta1998,giroletti2020}. 
The 86 and 230 GHz data were acquired under the POLAMI (Polarimetric Monitoring of AGN at Millimetre Wavelengths) Program\footnote{http://polami.iaa.es} \citep{agudo2018a,agudo2018b,thum2018}. The reduction and calibration of the POLAMI data are described in detail in \citet{agudo2010,agudo2014,agudo2018a}.
Fig.~\ref{radop} compares the radio flux densities with the composite $R$-band flux-density optical light curve. The optical magnitudes have been converted into de-reddened flux densities as explained in Section~\ref{sec_col}.

The radio flux density tends to increase with frequency, implying an inverted spectrum, with turning point around 86 GHz. 
Average values are 1.0, 1.1, 1.6, 1.6, and 1.2 Jy at 4.8, 8.5, 37, 86, and 230 GHz, respectively.
In the same bands, the variability amplitude $(F_{\rm max}-F_{\rm min})/F_{\rm mean}$ is 41\%, 64\%, 83\%, 35\% and 49\%. 
By considering that very uncertain data may artificially increase flux variability, we can rather calculate the fractional variability $f_{\rm var}=\sqrt{(\sigma^2-\delta^2)}/F_{\rm mean}$, where $\sigma^2$ is the flux variance and $\delta^2$ is the mean square error \cite[e.g.][]{vaughan2003}. We then obtain 10\%, 12\%, 22\%, 11\%, 19\% at 4.8, 8.5, 37, 86, and 230 GHz, respectively. In both cases, flux variations result much stronger at 37 GHz, which however is the best-sampled light curve. For comparison, the fractional variability of the $R$-band flux densities is $\sim 47\%$.

In the 37 GHz light curve we notice a large dip around 2020.1, at the same time of the, more rapid, optical dimming.
Whether the radio and optical events are physically linked or not is difficult to assess.

\begin{figure}
	\includegraphics[width=\columnwidth]{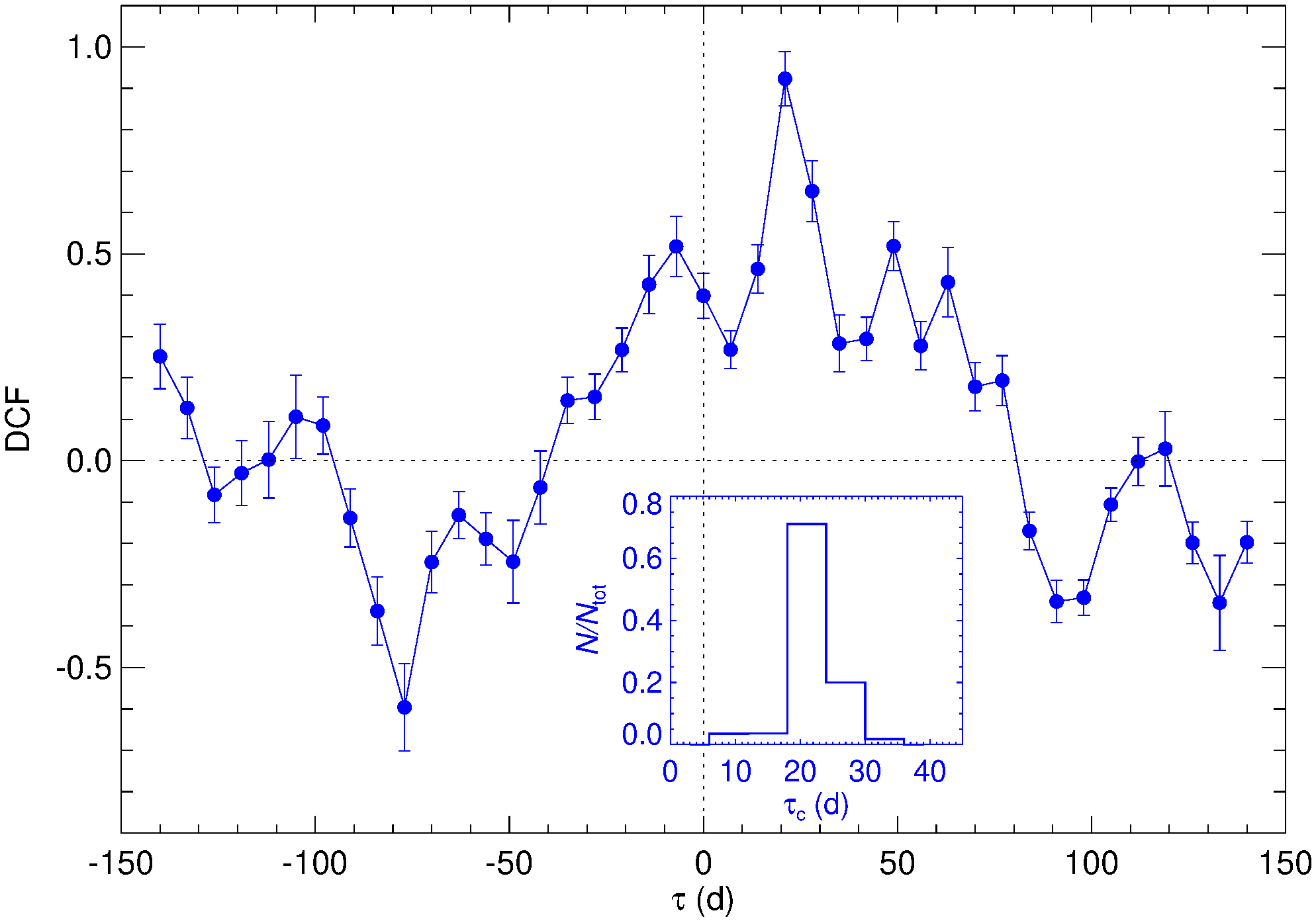}
    \caption{DCF between the composite WEBT + {\it TESS} flux-density light curve and the radio flux densities at 37 GHz. Optical data have been preliminarily averaged over 12 h bins, while the DCF has a 7 d resolution. The inset shows the distribution of the centroids of the DCF peaks obtained through 2000 FR/RSS Monte Carlo simulations.
}
    \label{dcf_or}
\end{figure}

One method that is commonly used to study the correlation between two unevenly-sampled time series is the discrete-correlation function \citep[DCF][]{edelson1988,hufnagel1992}. Peaks in the DCF represent correlations, whose significance increases as the DCF value of the peaks approaches or even overcomes unity. The DCF cross-correlation between the composite WEBT and {\it TESS} flux-density light curve of S4~0954+65 and its 37 GHz flux densities is shown in Fig.~\ref{dcf_or}. The main peak occurs at a time lag $\tau = 21 \rm \, d$, which indicates a delay of about three weeks of the 37 GHz radio variations after the optical ones. To check the uncertainty on this time delay, we ran 2000 flux randomization/random subset selection (FR/RSS) Monte Carlo simulations \citep{peterson1998,raiteri2003}. For each of the resulting DCFs, we calculated the centroid of the peak. Fig.~\ref{dcf_or} shows the centroid distribution: 71\% ($\ga 1 \sigma$) of the simulations resulted in a time lag lying in the interval 18--24 d. 
A radio delay of several days or weeks is often found in blazars and has been interpreted as an indication that the radio emission comes from a jet region which is located downstream of the optical emission zone \citep[e.g.][]{raiteri2017_nature}.

\section{Polarization}
\label{sec_pola}
Blazar synchrotron emission is polarized and 
both the degree of optical polarization $P$ and the electric vector polarization angle EVPA are extremely variable \citep[e.g.][]{smith1996,marscher2014}. In particular, wide rotations of the EVPA are often observed
\citep[e.g.][]{marscher2008,blinov2015,blinov2016,blinov2018,hagen2015,raiteri2017,gupta2017,lyutikov2017,zhang2020}
 \begin{figure}
	\includegraphics[width=\columnwidth]{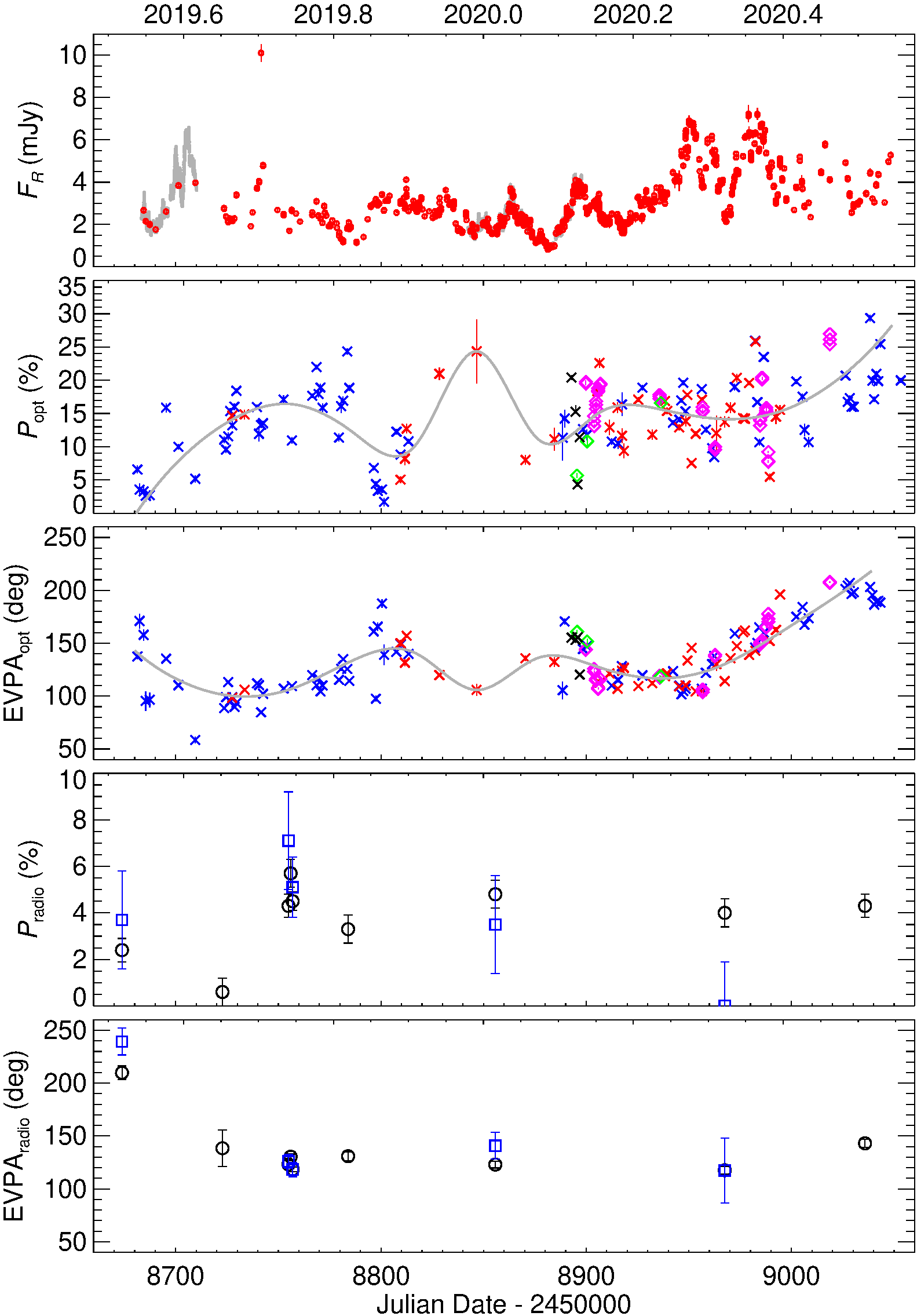}
    \caption{From top to bottom: a) De-reddened WEBT (red dots) and {\it TESS} (grey dots) flux densities in the $R$ band.
b) Degree of optical polarization. 
c) Optical EVPA.
d) Degree of radio polarization at mm wavelengths.
e) Radio EVPA.
In the last four panels, the various datasets are distinguished with different symbols and colours, according to Table~\ref{obs}. 
Grey solid lines in panels b) and c) represent cubic spline interpolations on the 30-d binned data to better follow the general trend.
}
    \label{pola}
\end{figure}

\begin{figure}
	\includegraphics[width=\columnwidth]{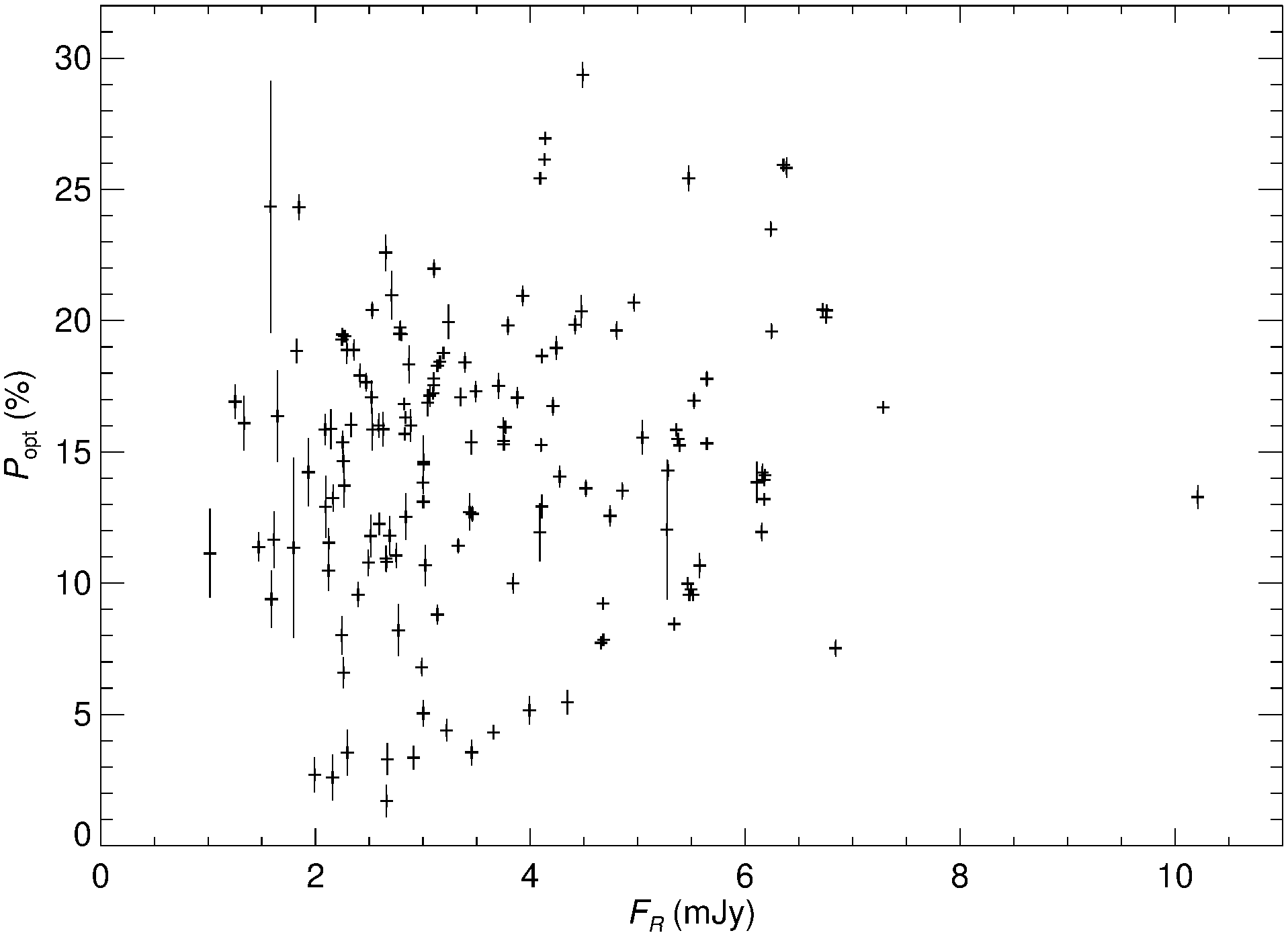}
    \caption{The degree of optical polarization $P_{\rm opt}$ versus the de-reddened flux density in the $R$ band, $F_R$, showing that there is no clear correlation between these two variables.
}
    \label{pf}
\end{figure}

During the time span covered in this paper, polarization observations of S4~0954+65 in the optical band were done at the Crimean (70 and 260 cm telescopes), Perkins, San Pedro Martir, and St.~Petersburg observatories.
The results are shown in Fig.~\ref{pola} and compared with the $R$-band composite light curve built with WEBT and {\it TESS} data.
The value of $P_{\rm opt}$ ranges between 1.7\% and about 29\% with continuous oscillations, which do not seem to be correlated with flux. This is more clearly displayed in Fig.~\ref{pf}, where $P_{\rm opt}$ is plotted versus the flux density.

The polarization angle has a more regular oscillating behaviour.
If we trace the average trend of both the polarization degree and angle with a cubic spline interpolation on the 30-d binned $P_{\rm opt}$ and $\rm EVPA_{opt}$, we can see that the $\rm EVPA_{opt}$ inverts its way of rotation when $P_{\rm opt}$ reaches a maximum or a minimum.
However, the irregular sampling does not allow any clear conclusion.

In Fig.~\ref{pola} we also plotted the radio polarization degree and angle at millimetre wavelengths. The values of $P_{\rm radio}$ and $\rm EVPA_{radio}$ at 230 and 86 GHz are mostly in agreement. 
In comparison to what happens in the optical band, the radio polarization degree and angle seem much less variable. However, this could be an effect of the sparser sampling.
The mean polarization angles are similar in the two bands, around 140\degr.

\section{Short-term variability analysis}
\label{sec_stv}
In this section we explore the short-term variability of the source using the exceptionally well-sampled {\it TESS} light curves to see whether characteristic variability time-scales can be identified.

\subsection{Autocorrelation Function}
\label{sec_acf}

When the DCF introduced in Section~\ref{sec_radio} is used to cross-correlate a light curve with itself, its auto-correlation function ACF is obtained. This is symmetric around the zero time lag, where it has a peak. 
Variability time-scales, i.e.\ distances between minima and maxima in the light curve, appear as minima in the ACF, while periodicities (or quasi-periodicities) will give rise to (approximately) equally spaced ACF maxima.

\begin{figure}
	\includegraphics[width=\columnwidth]{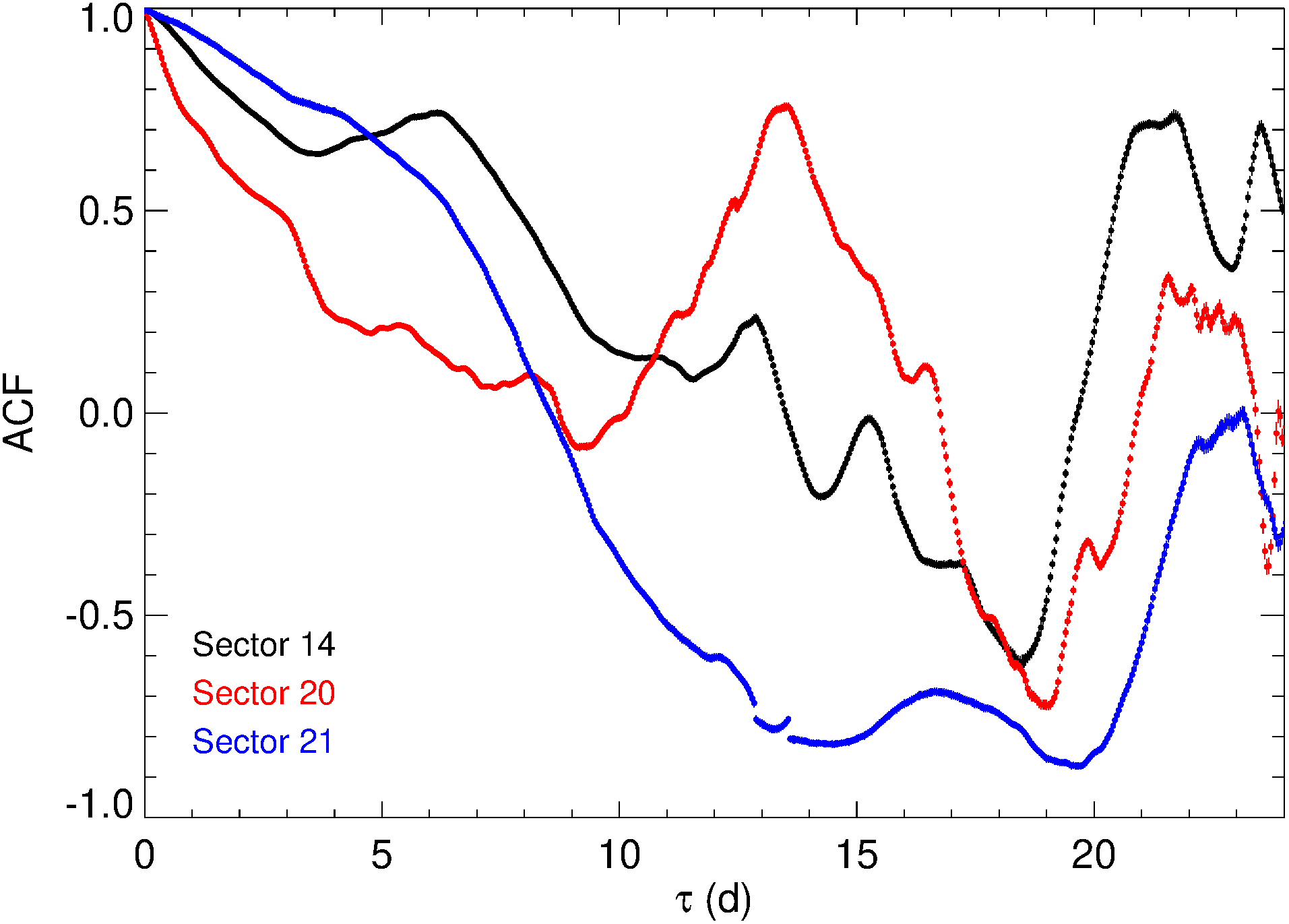}
    \caption{Auto-correlation function of the {\it TESS} light curve in Sector 14 (black), Sector 20 (red), and Sector 21 (blue). The ACF is binned in 1 h time lag intervals. 
}
\label{acf_all}
\end{figure}

The 2 min-cadence observations by {\it TESS} along three time spans of almost one month each allow us to analyse in detail the optical variability of S4~0954+65 on time-scales from about 1 h to a few weeks.
In Fig.~\ref{acf_all} we show the ACF for the {\it TESS} light curve in each of the observing sectors, binned in 1 h time lag intervals.
The behaviour appears quite different in the three sectors. Moreover, the signature of the shortest variability time-scales is overwhelmed by that of the long-term trend.

 \begin{figure}
	\includegraphics[width=\columnwidth]{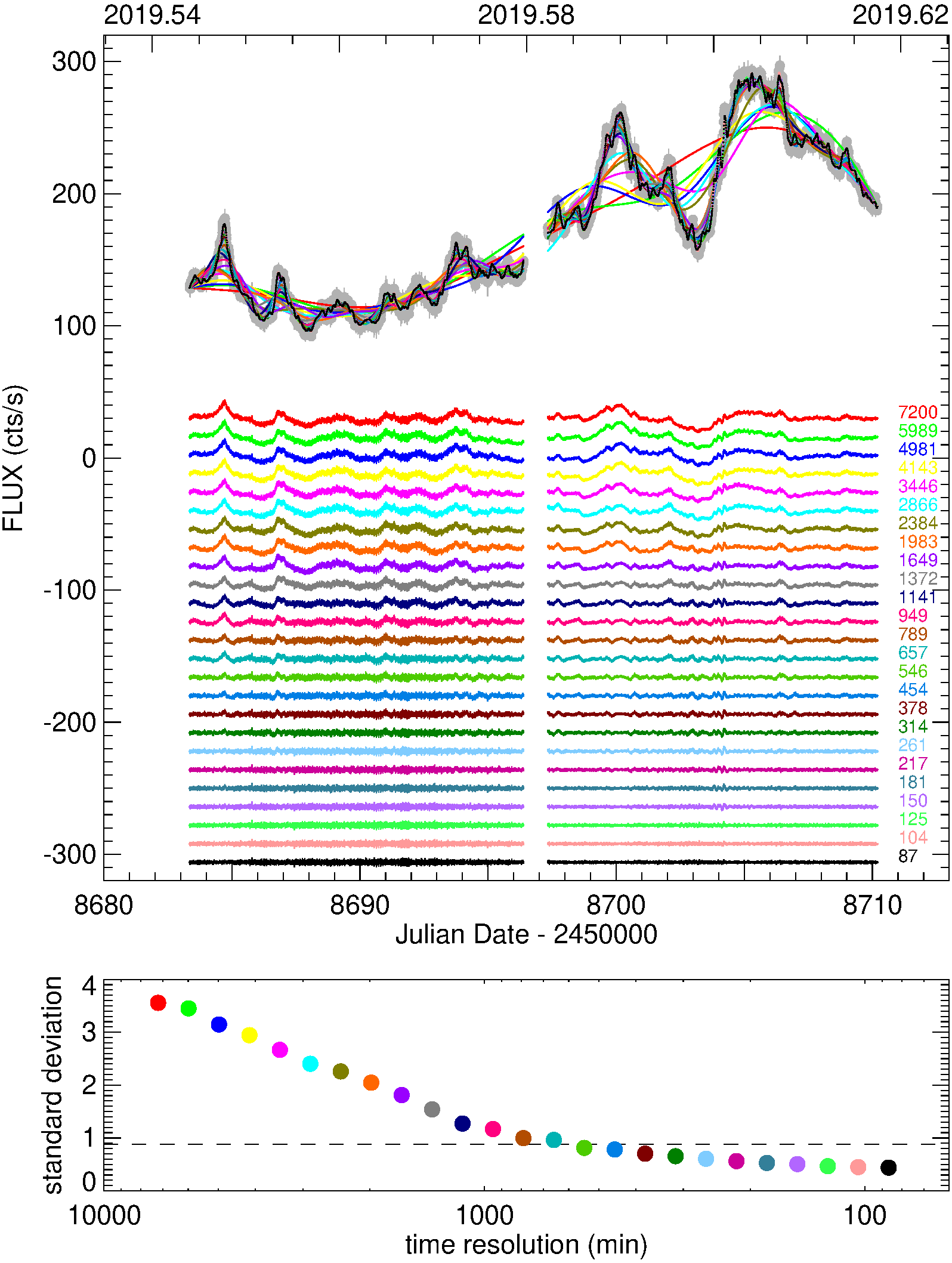}
    \caption{The top panel displays the {\it TESS} light curve (grey) in Sector 14. Coloured cubic spline interpolations, representing long-term trends with different time resolution between 7200 min (red) to 87 min (black), are superimposed to it. The detrended light curves obtained by dividing the {\it TESS} light curve by the splines are shown with the same colours, highlighting how the variability amplitude decreases with increasing long-term trend time resolution. This is better shown in the bottom panel, where the standard deviation of the detrended light curve is plotted versus the long-term resolution, i.e.\ the time bin used to obtain the long-term trend through cubic spline interpolation of the {\it TESS} binned light curve.
The dashed horizontal line marks the mean error of the detrended light curves.
}
    \label{detre14}
\end{figure}

 \begin{figure}
	\includegraphics[width=\columnwidth]{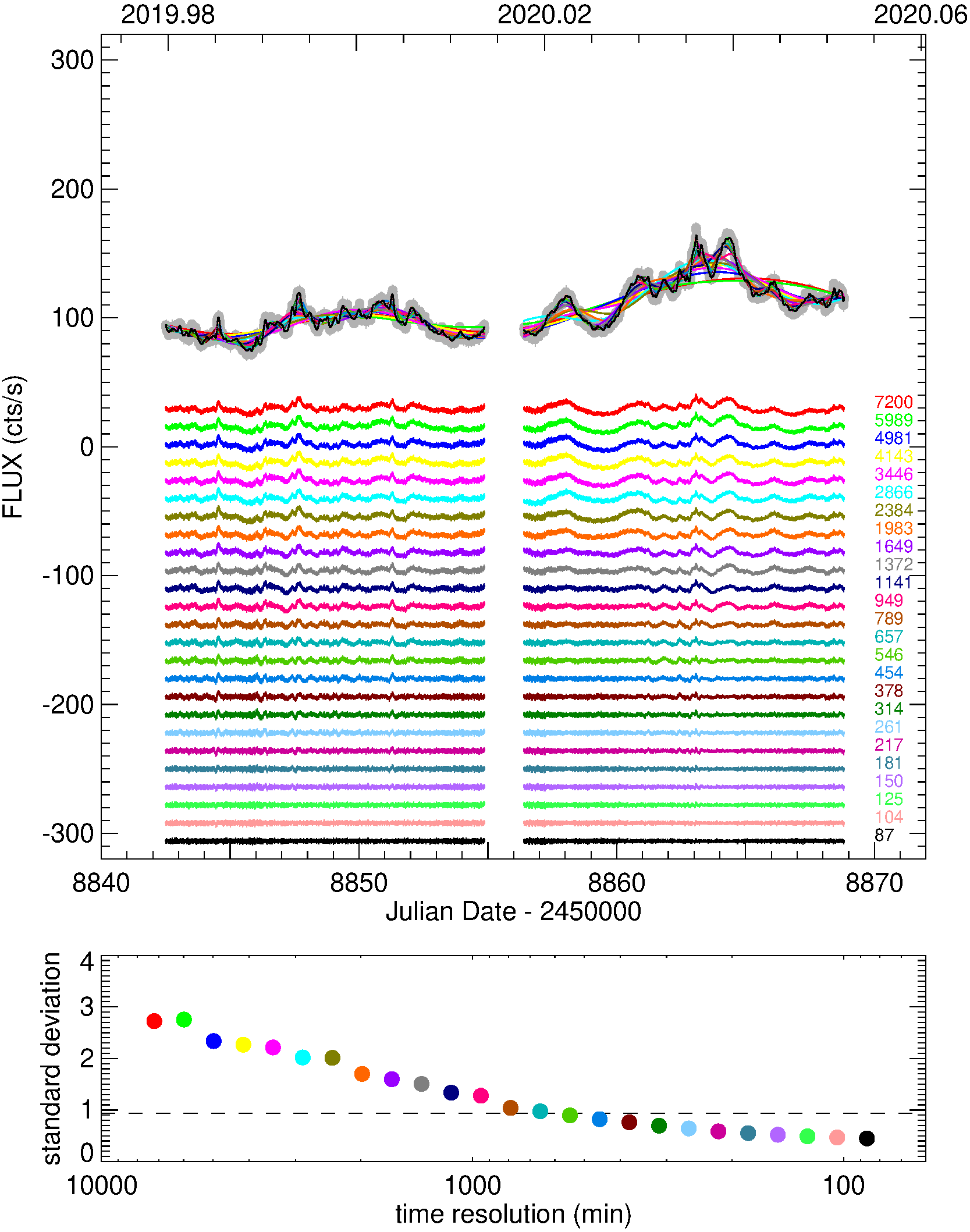}
    \caption{As in Fig.~\ref{detre14}, but for Sector 20.}
    \label{detre20}
\end{figure}

 \begin{figure}
	\includegraphics[width=\columnwidth]{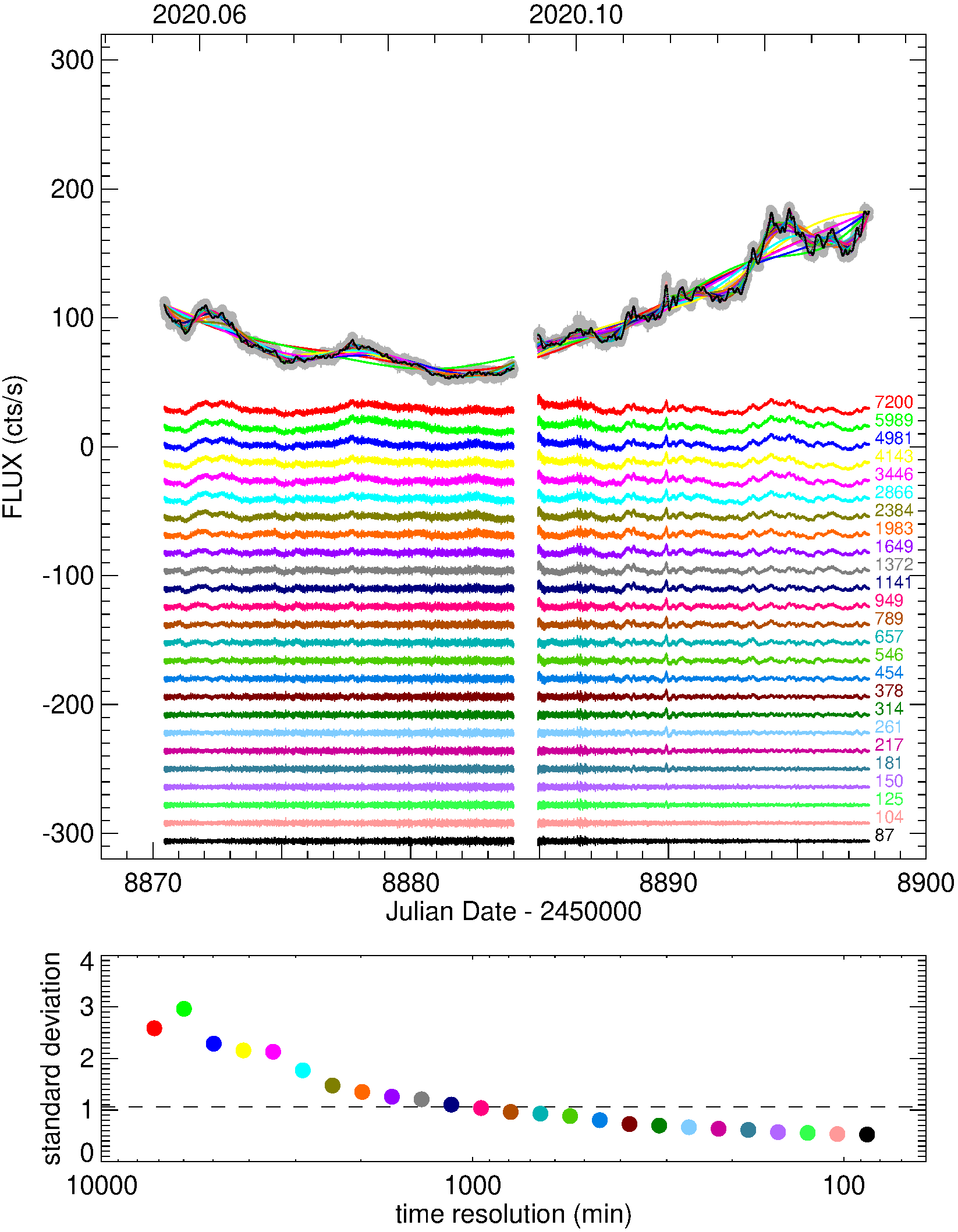}
    \caption{As in Fig.~\ref{detre14}, but for Sector 21.}
    \label{detre21}
\end{figure}

Therefore, to analyse the short-term variability of the {\it TESS} fluxes, we apply the detrending method developed by \citet{raiteri2021}.
The method originates from the principle that in order to unveil the shortest variability time-scales one needs to correct for the long-term trend first.
Because the boundary between long- and short-term variability is somewhat undefined,  
the idea is to correct the light curve for long-term trends with progressively higher time resolution to see what variability time-scales can be uncovered.
Following this line, we first bin separately the Sector 14, 20 and 21 light curves, using for each of them time bins going from 7200 min (5 d) to 87 min, in 25 steps equally spaced in logarithm. 
We then interpolate cubic splines through the 25 binned light curves of each sector. These splines represent long-term trends with different time resolutions. 
We stress that when we speak of the time resolution of a given long-term trend, we mean the time interval used to bin the {\it TESS} light curve whose cubic spline interpolation represents the long-term trend itself.

By dividing the {\it TESS} fluxes by the splines (and multiplying by a constant value), detrended light curves are finally obtained, 25 for each sector.
Figs.~\ref{detre14}--\ref{detre21} show the {\it TESS} fluxes in each of the three sectors, together with their 25 cubic spline interpolations representing long-term trends with progressively higher time resolution, and the corresponding detrended light curves, properly scaled.
These display a variation amplitude that decreases with increasing time resolution of the long-term trend, i.e.\ as the spline enters the bends of the {\it TESS} light curve more deeply and thus reproduces it in more details. 
The figure also quantifies this by showing the standard deviation of the detrended light curves versus the time resolution of the long-term trend used to obtain them. The mean error of the detrended light curves is 0.9 cts/s in Sector 14 and 20, and it is 1.0 cts/s in Sector 21.
Detrended light curves obtained from time bins shorter than about 600 min in Sector 14 and 20, and about 1000 min in Sector 21, have standard deviation smaller than the mean error. However, well-defined flux variations can still be recognized in these detrended light curves, suggesting that they are real flux changes.

 \begin{figure}
	\includegraphics[width=\columnwidth]{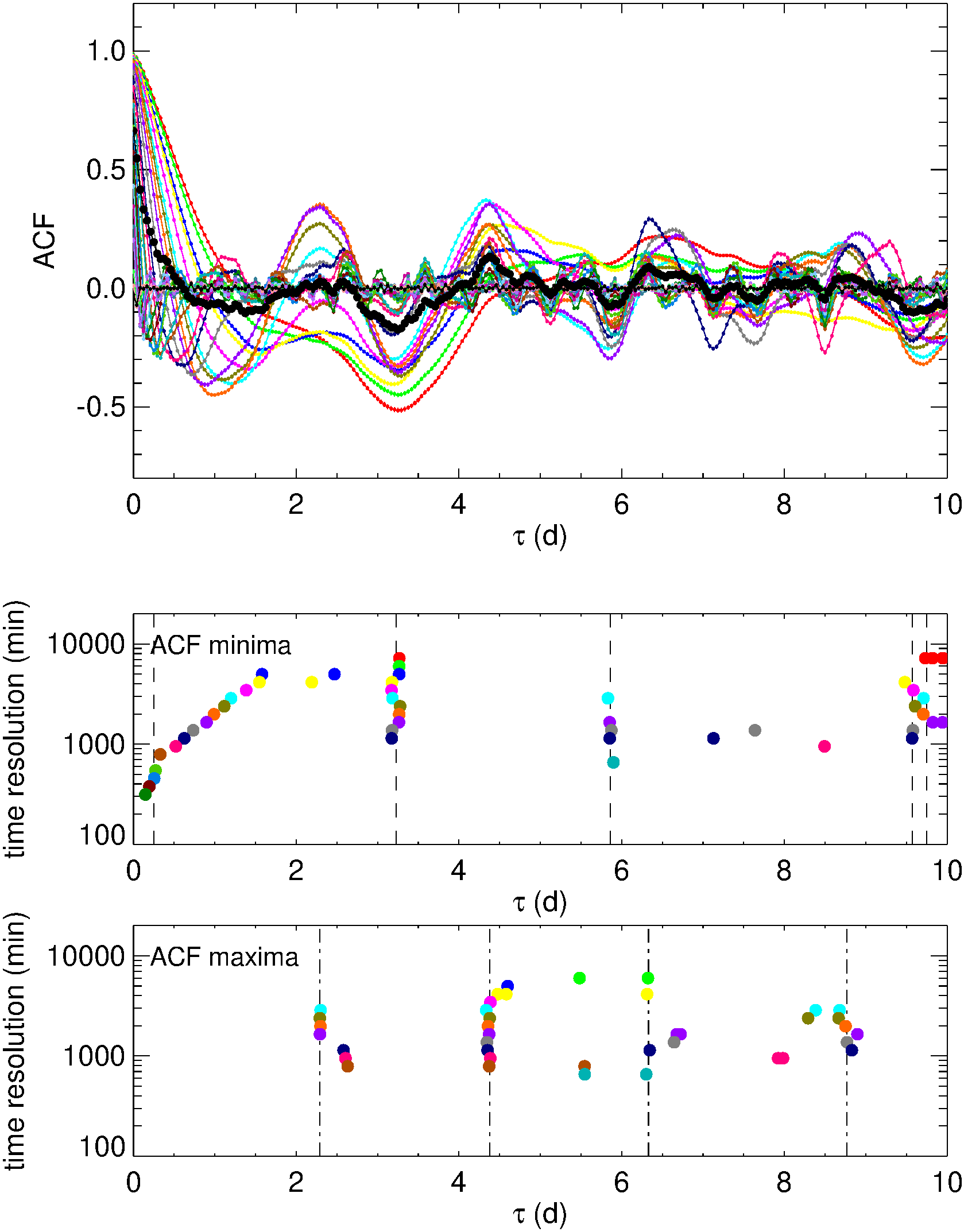}
    \caption{Top panel: the ACFs of the detrended light curves of Sector 14 shown in Fig.~\ref{detre14}. The black dots represent the average ACF, whose minimum and second maximum are used as thresholds for the significance of the minima/maxima of the single ACFs. Middle/Bottom panel: the time lag of the most significant ACF minima/maxima versus the time resolution of the spline used to perform the detrending. Vertical lines mark the results of the weighted means on the most significant signals. They are also reported in Table~\ref{ts}. In the bottom panel, we do not show the absolute maxima at the origin.}
    \label{acf14}
\end{figure}

 \begin{figure}
	\includegraphics[width=\columnwidth]{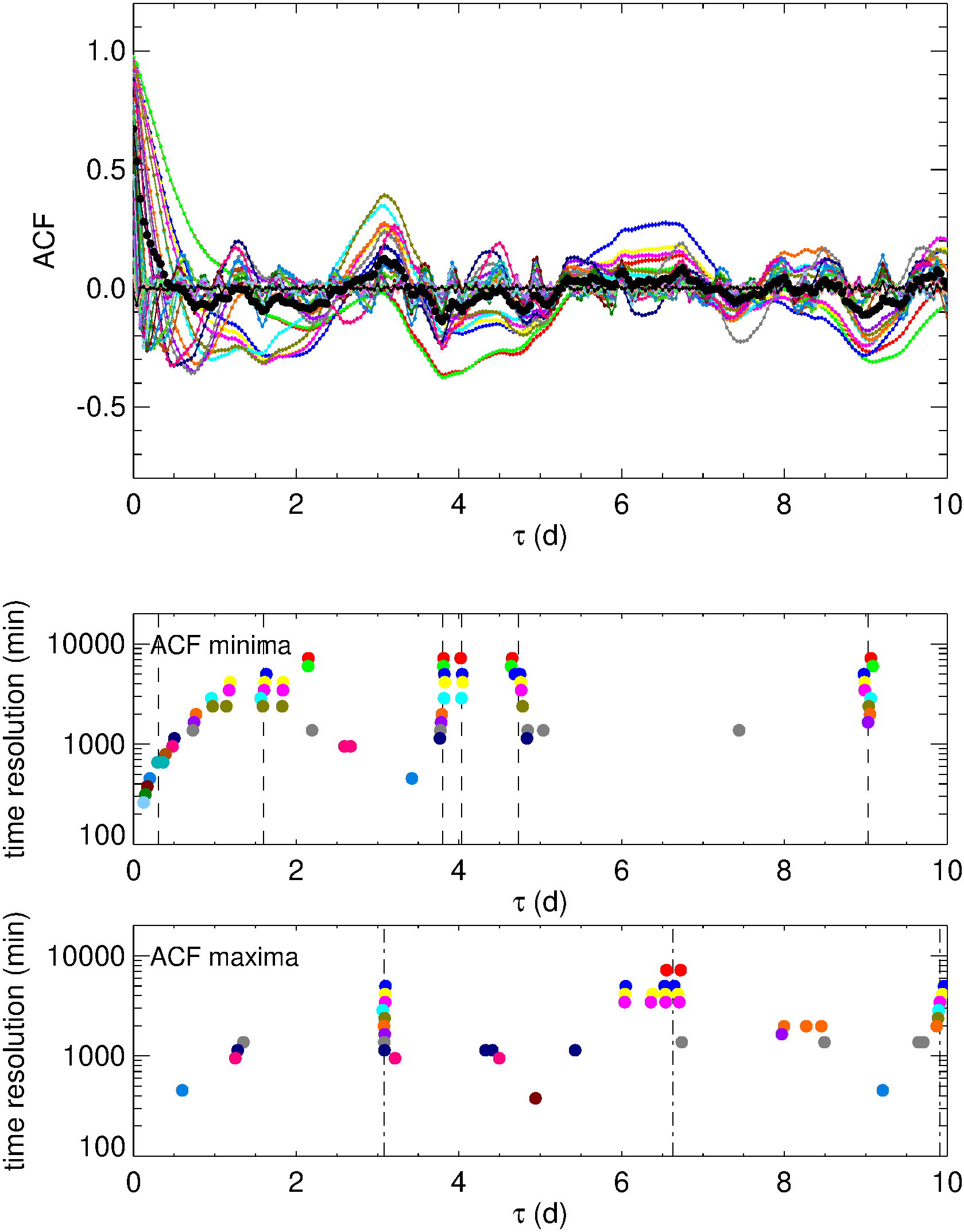}
    \caption{As in Fig.~\ref{acf14}, but for Sector 20.}
    \label{acf20}
\end{figure}
 
\begin{figure}
	\includegraphics[width=\columnwidth]{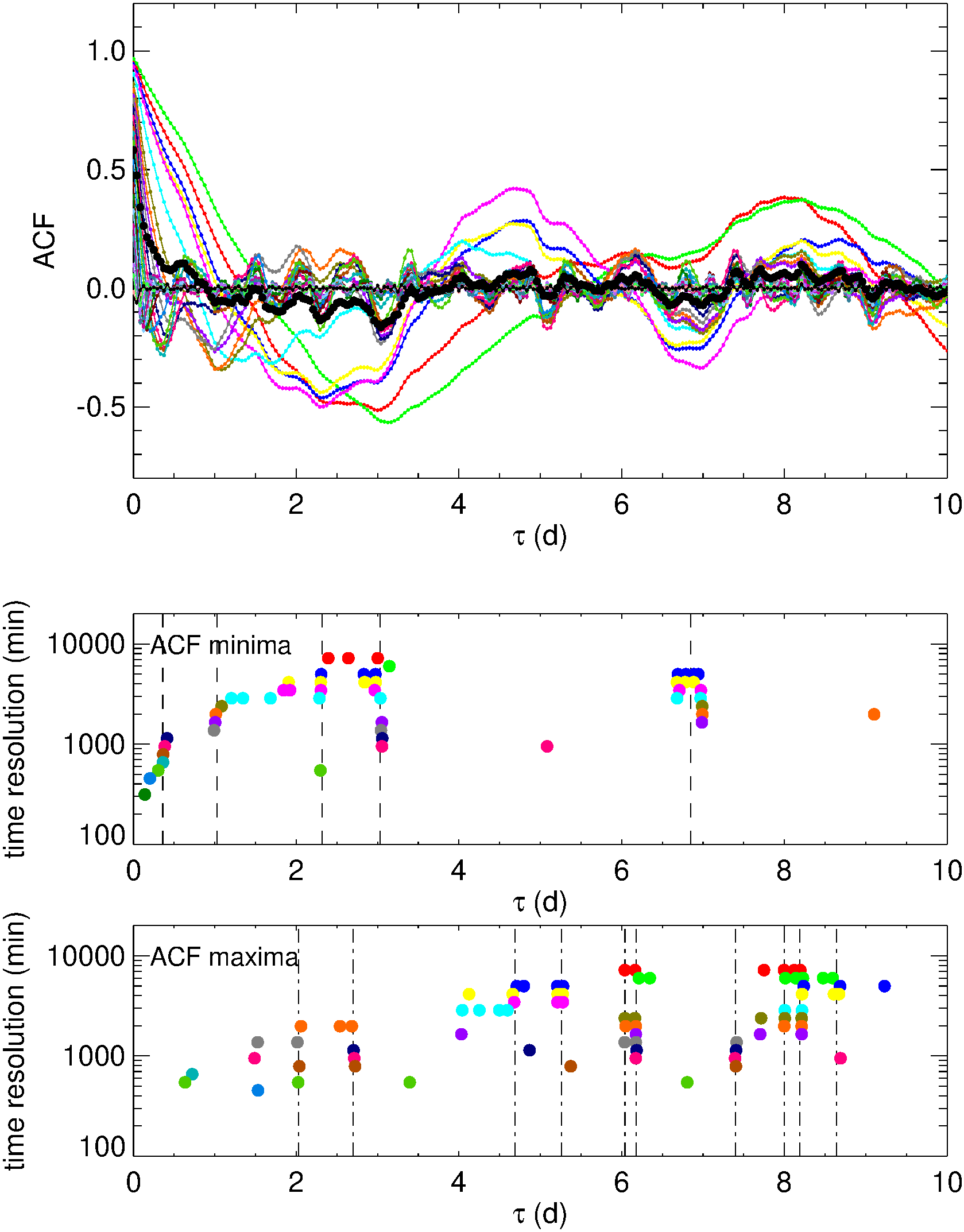}
    \caption{As in Fig.~\ref{acf14}, but for Sector 21.}
    \label{acf21}
\end{figure}

The ACFs obtained from the detrended light curves in each sector are shown in Figs.~\ref{acf14}--\ref{acf21} for time lags shorter than 10 d. 
For each sector, ACF minima are visible at various time lags. However, some of them concentrate at approximately the same $\tau$, meaning that they do not depend on the spline resolution used to remove the long-term trend. 
Therefore, these minima can be considered reliable variability time-scales of the source. 
Similarly, there are ACF maxima at roughly the same $\tau$, which may indicate possible quasi-periodicities.
To identify the most significant signals, we found the position of the minima and maxima of each ACF by imposing limits to the first and second derivatives. 
Moreover, for each ACF we only considered those minima/maxima whose ACF values are smaller/larger than certain thresholds.
These were fixed by calculating an average ACF and using its minimum as the threshold for the single ACF minima, and its second maximum (besides that in the origin) as a threshold for the single ACF maxima. The thresholds for the minima were thus set to -0.175, -0.139, and -0.164 for Sector 14, 20, and 21, respectively. The corresponding thresholds for the maxima were 0.133, 0.125, and 0.107.
We then performed a weighted mean of the closest signals, with weights proportional to their ACF values and finally selected the minima and maxima that show up more frequently in the ACFs of a given sector, i.e.\ with more than 3 elements contributing to the mean.
The time lags of the  minima represent the most significant variability time-scales, while those of the roughly equally spaced maxima represent possible quasi-periodicities. The time lags of the significant minima and maxima are reported in Table~\ref{ts} for each sector.

\begin{table*}
	\centering
	\caption{Results of the time-series analysis on the {\it TESS} light curves in Sectors 14, 20, 21, and the whole composite optical light curve.
The tabulated $\tau$ values derived from the ACF on the detrended {\it TESS} light curves are obtained as weighted means of the most recurrent signals; the numbers in brackets indicate the number of elements contributing to the corresponding weighted mean.}
	\label{ts}
	\begin{tabular}{l|l}
\hline
\multicolumn{2}{c}{\it TESS light curves}\\
\hline
Sector & $\tau$ of ACF minima (d) \\                                 
14 & 0.25 (5), 3.23 (11), 5.86 (5),  9.57 (5), 9.75 (5)\\
20 & 0.31 (9), 1.60 (5), 3.80 (9), 4.03 (4), 4.73 (9), 9.03 (9) \\
21 & 0.36 (5), 1.03 (4), 2.32 (6), 3.03 (10), 6.85 (14)\\
\hline
Sector & $\tau$ of ACF maxima (d)\\ 
14 & 2.29 (4), 4.38 (10), 6.33 (4), 8.77 (6) \\
20 & 3.08 (9), 6.63 (9), 9.91 (6) \\
21 & 2.03 (4), 2.70 (4), 4.69 (7), 5.26 (7), 6.04 (4), 6.18 (8), 7.40 (4), 8.00 (5), 8.19 (10), 8.64 (5) \\
\hline
Sector & PSD peaks (d) with period $\geq 0.5 \, \rm d$\\ 
14     & 0.90 \\
20     & 0.62, 1.10, 1.39 \\
21     & -- \\
\hline
\multicolumn{2}{c}{\it Composite light curve}\\
\hline
$\tau$ of ACF minima (d) & 18, 73, 105, 139, 173 \\
$\tau$ of ACF maxima (d) & 33, 91, 154, 190--193 \\
PSD peaks (d)  & 31, 39.7\\

\hline
\end{tabular}
\end{table*}

\begin{figure}
	\includegraphics[width=\columnwidth]{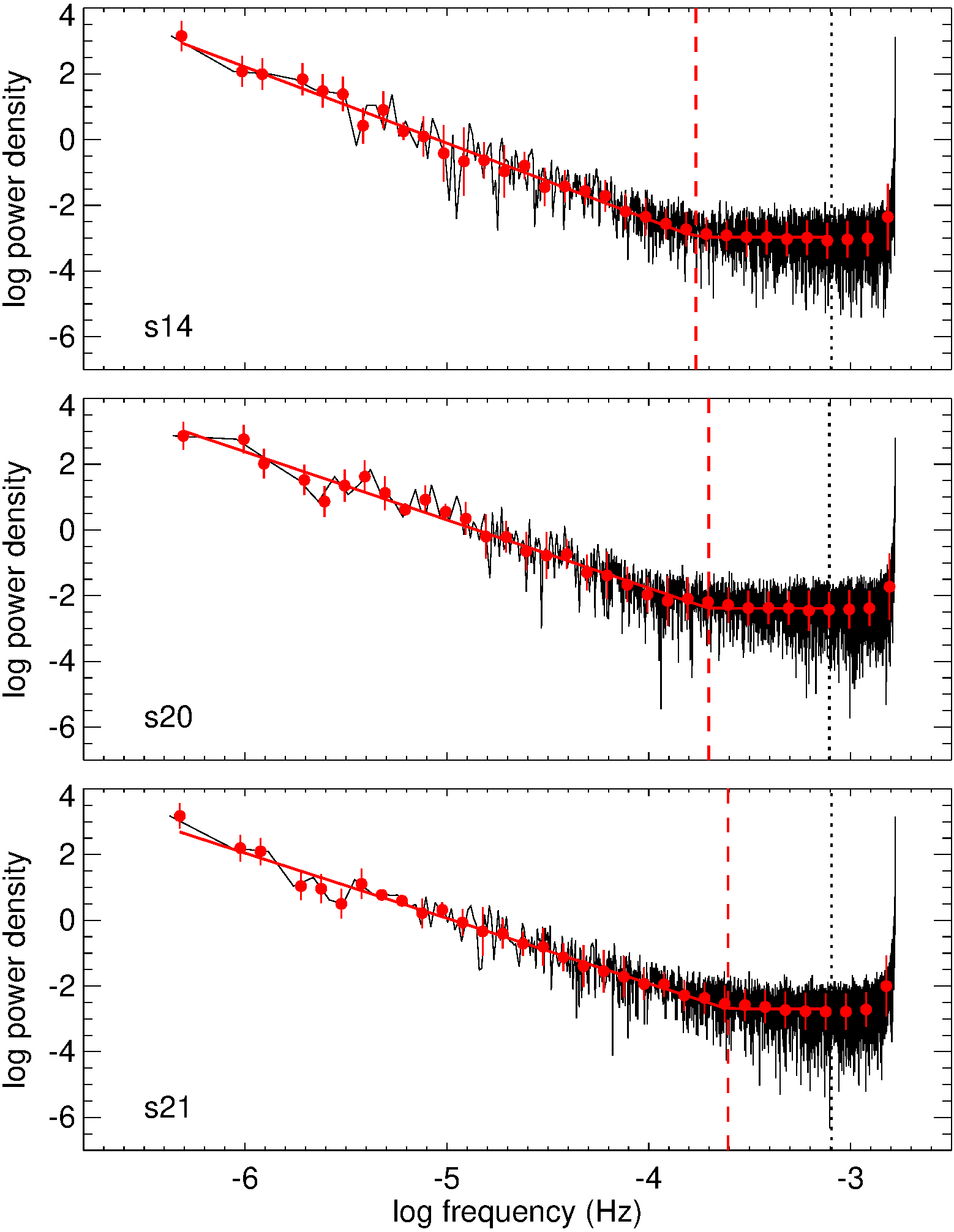}
    \caption{Frequency-sampled PSD computed on the {\it TESS} light curves in the three sectors. 
The dotted vertical lines mark the Nyquist frequency. The solid red lines represent a broken power-law fit to the PSD, binned in steps of $\log f = 0.1$ up to the Nyquist frequency. The vertical dashed red lines indicate the best-fit break frequency.}
    \label{tisean_fre}
\end{figure}

\begin{figure}
	\includegraphics[width=\columnwidth]{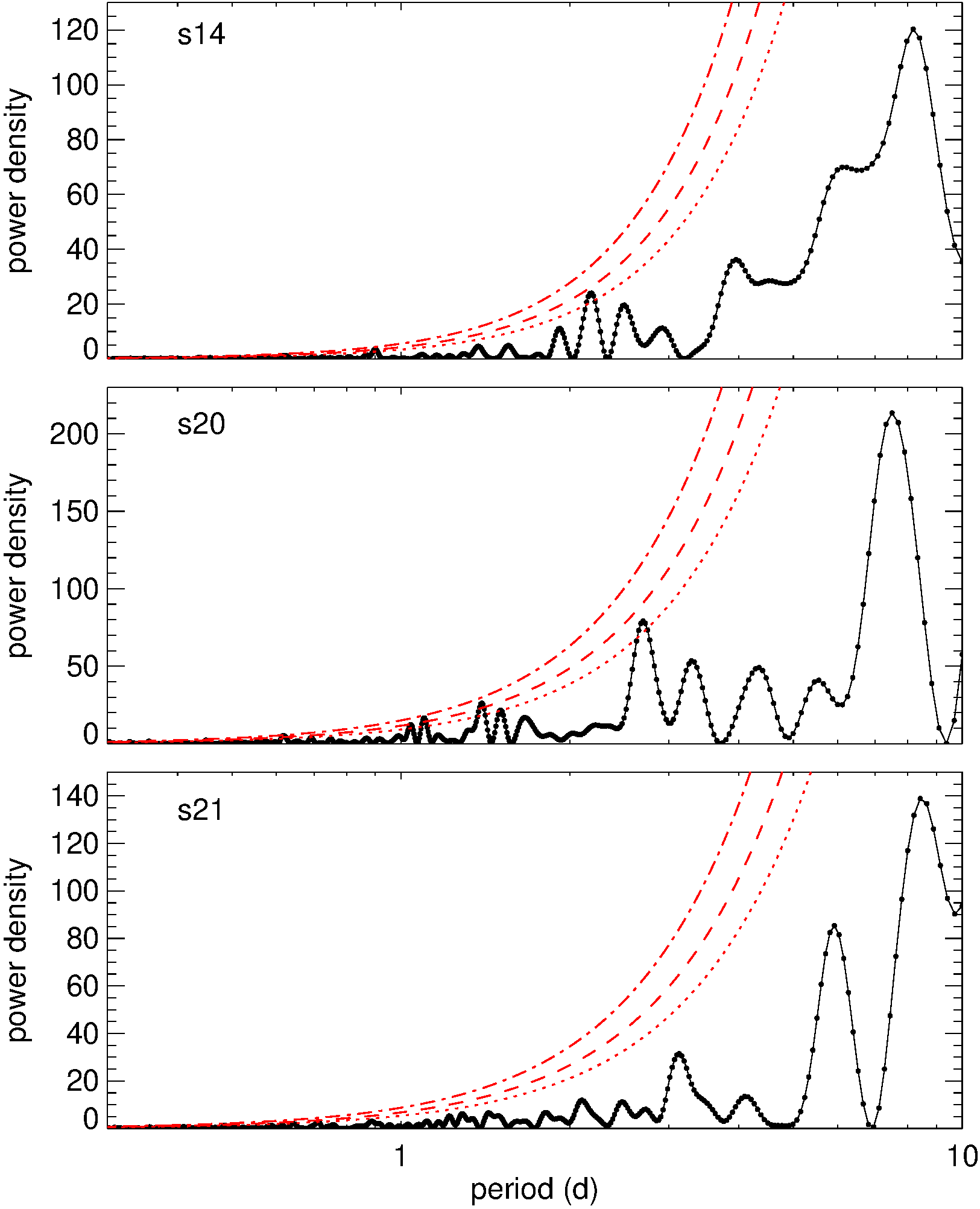}
    \caption{Time-sampled periodogram computed on the {\it TESS} light curves in the three sectors. Dotted, dashed, and dot-dashed red lines indicate the significance levels of 95, 99, and 99.9 per cent, respectively.}
    \label{tisean_per}
\end{figure}

\begin{figure}
	\includegraphics[width=\columnwidth]{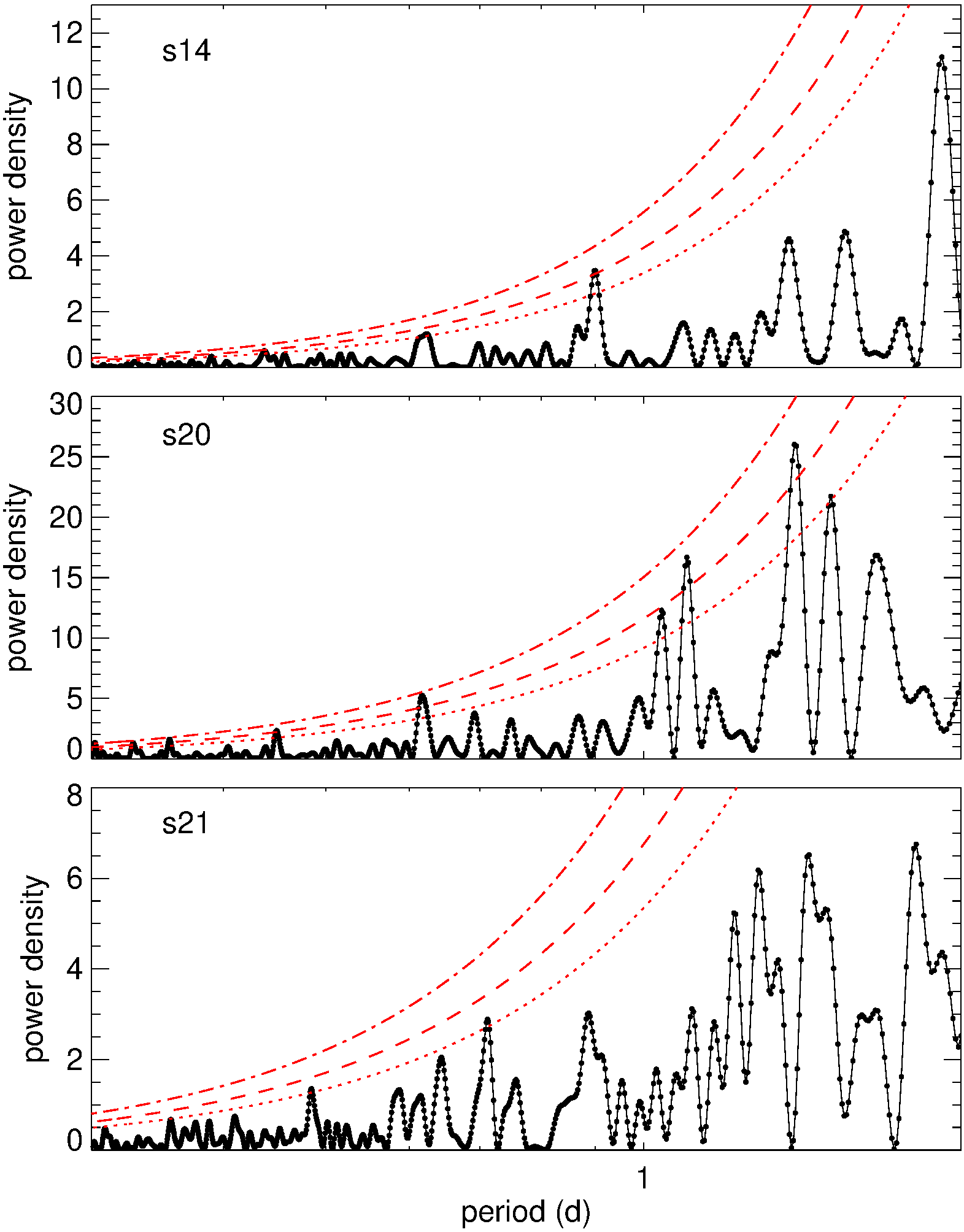}
    \caption{A zoom of Fig.~\ref{tisean_per} into the 0.3--2 d period range.}
    \label{tisean_per_zoom}
\end{figure}

We adopted the same detrending method, but using the structure function SF \citep{simonetti1985} instead of the ACF, to check the results reported in Table~\ref{ts}. This analysis is detailed in the Appendix and confirms the above findings, with the addition of possible further characteristic variability time-scales.

We note the interesting sequence of ACF maxima at 2.3, 4.4, 6.3, and 8.8 d in Sector 14, the sequence 3.1, 6.6, and 9.9 d in Sector 20, and the sequence 2.7, 5.3, and 8.0 d in Sector 21. These sequences may indicate quasi-periodicities, which will be further investigated with the periodogram in the next section.

In any case, the source short-term variability appears strongly unstable, with multiple time-scales from a few hours to a few days, which change from time to time.

\subsection{Periodogram}
\label{sec_per}
As mentioned earlier, the nearly equally spaced ACF significant maxima derived in Section~\ref{sec_acf} and whose time lags are reported in Table~\ref{ts} may indicate possible periodic patterns in the source variability. The reliability of them can be checked with the periodogram, which is an estimate of the power spectral density (PSD)\footnote{In the following, we will use the term ``PSD" as a synonym of ``periodogram", as is commonly used.}.
Fig.~\ref{tisean_fre} shows the PSD of the {\it TESS} light curves in the three sectors, obtained with the normalized Lomb-Scargle periodogram \citep[][see also \citealt{raiteri2021}]{scargle1982,horne1986}.
They have been computed in the frequency range $\sim 4 \times 10^{-7}$ to $\sim 1.67 \times 10^{-3} \, \rm Hz$,
the first limit corresponding to the duration of the light curve and the second to ten minutes, which is the width of the light curve bins (see Section~\ref{sec_tess}). Indeed, at this frequency the PSDs show a clear spike. 
Because of the frequency sampling, the PSDs are better sampled at the highest frequencies, where they show a flattening, i.e.\ a white noise behaviour.
We thus binned the PSDs in steps of $\log f = 0.1$ [Hz] and used linear regression with $\chi^2$ error statistic minimization to fit them with a broken power law up to the Nyquist frequency\footnote{The Nyquist critical frequency is defined as $f_c=N/(2 \, \Delta T)$, where $N$ is the number of data points in the time interval $\Delta T$.}, with slope fixed to zero after the break frequency.
The breaks occur at $\log f =-3.77 \pm 0.24, -3.70 \pm 0.27, -3.61 \pm 0.30$  [Hz] for Sector 14, 20 and 21, respectively (corresponding to $\sim 5888$,  5012 and 4074 s), while the slopes of the power law before the break are $-2.32 \pm 0.30, -2.07 \pm 0.28, -1.98 \pm 0.26$, with significance levels of the Kolmogorov-Smirnov statistic of 0.936 in Sectors 14 and 20, and 0.997 in Sector 21.
This confirms the red noise nature of the source variability up to frequencies corresponding approximately to 1--1.5 h, while for higher frequencies white noise dominates.

PSDs sampled in time for the three sectors are displayed in Fig.~\ref{tisean_per}.
Significance levels of 95, 99 and 99.9 per cent are calculated as in \citet{raiteri2021}, following the prescriptions given by \citet{vaughan2005}. A zoom into the 0.3--2 d period range is shown in Fig.~\ref{tisean_per_zoom}.
None of the PSD peaks exceeds the 99.9 per cent significance level, while there are a few signals in Sectors~14 and 20 that overcome the 99 per cent level. These are reported in Table~\ref{ts}, where they can be compared with the results obtained from the time-series analysis by means of the ACF. 
The comparison shows that there are no common values, i.e.\ there is no repeated time distance between the light curve peaks or dips that correspond to the period of sinusoidal components.
The reason may be that we are in the presence of quasi-periodic signals, so that the power density is spread over different times, becoming weaker \citep{vaughan2005}. In any case, we cannot conclude in favour of short-term periodicities in the S4~0954+65 optical light curves.

We finally note that the marginally significant signal at 0.62 d present in Sector~20 is also visible in the ACF maxima of Sectors~20 and 21 (see Figs.~\ref{acf20} and \ref{acf21}) and in the SF minima of Sector 20 (see Fig.~\ref{sf20}), thus representing an interesting recurrence. Moreover, in all sectors there is a significant ACF minimum at a time lag about one half of this signal (and similar values are found in the SF maxima in the Appendix).

\section{Long-term variability analysis}
\label{sec_ltv}

\begin{figure*}
	\includegraphics[width=12cm]{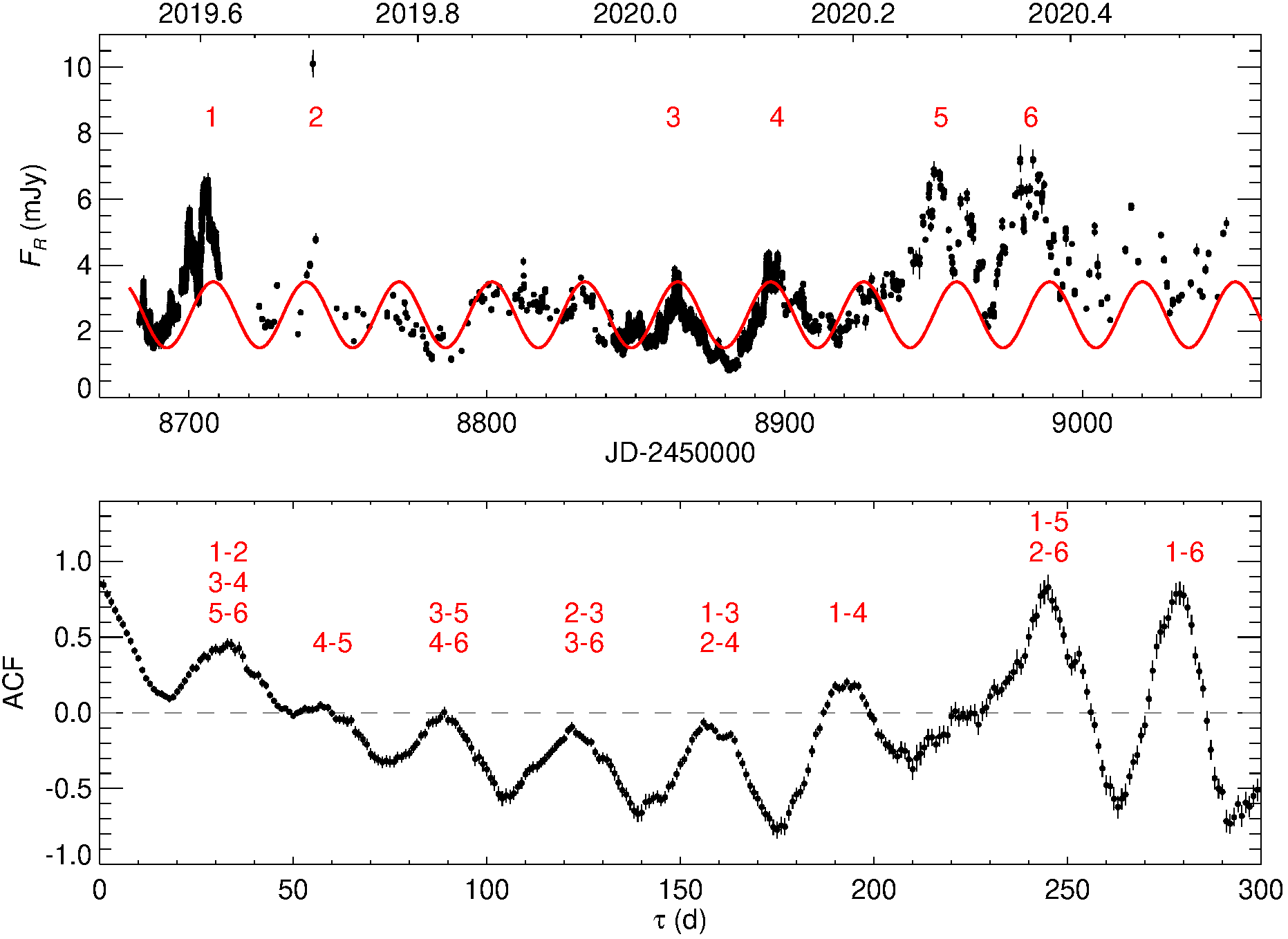}
    \caption{Top: black dots represent the composite light curve obtained by including both the WEBT data and the {\it TESS} data stretched to match the WEBT ones. Red numbers mark the main flares. The red solid line is a sinusoidal function with period of 31.2 d, as found by the PSD analysis.
Bottom: the ACF on the composite light curve (black).
Data points have preliminary been binned over 6 h, while the ACF resolution is 1 d. 
The red labels over the ACF maxima indicate the pairs of light curve peaks that are mainly responsible for them.}
    \label{acftw}
\end{figure*}

To investigate the presence of characteristic variability time-scales during the whole period analysed in this paper, we use the composite light curve including the WEBT data together with the {\it TESS} data stretched to match the WEBT ones (see Section~\ref{sec_webt}).
This is shown again in Fig.~\ref{acftw}, where the main flares have been numbered. They come in pairs, whose peaks are separated by an approximately one month time interval. This is true for the pairs 1-2, 3-4, and 5-6.

The ACF of the composite light curve is plotted in Fig.~\ref{acftw} and shows many signals, corresponding to the cross-correlation between different light curve peaks (and dips). In particular, the first maximum at about one month corresponds to the coupling of the 1-2, 3-4, and 5-6 pairs. 
The positions of the subsequent ACF maxima are roughly multiples of this time interval. The signals are produced by the pairs indicated in the figure.
This suggests the presence of a quasi-periodicity of roughly one month.

\begin{figure}
	\includegraphics[width=\columnwidth]{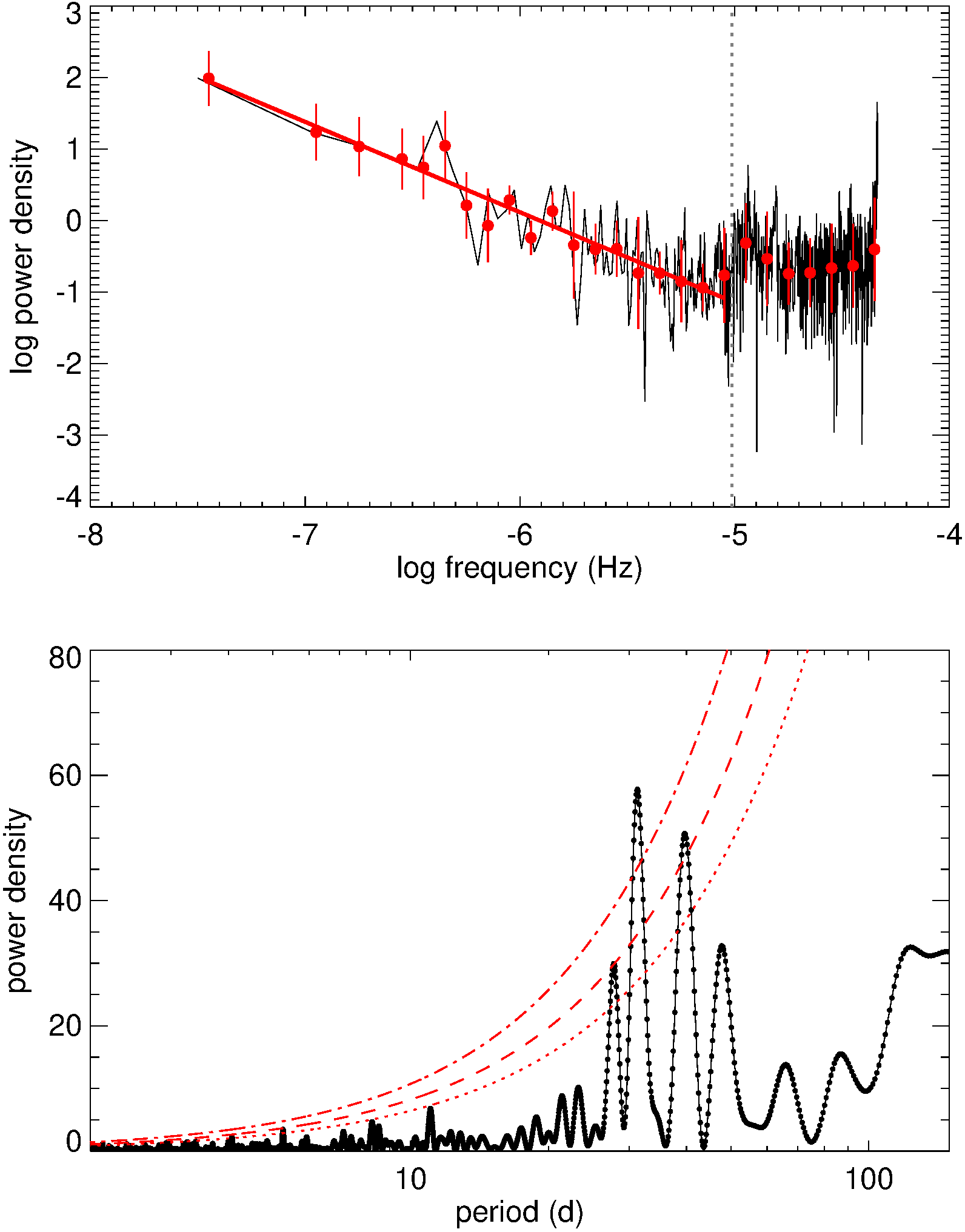}
    \caption{Top: Periodogram of the composite WEBT+{\it TESS} 6-h binned light curve sampled in frequency. Bottom: Periodogram of the same composite light curve sampled in period. }
    \label{tisean_tot}
\end{figure}

The result of the periodogram analysis on the composite light curve is shown in Fig.~\ref{tisean_tot} and confirms the quasi-periodicity of about one month found with the ACF. Indeed, a strong signal with significance level greater than 99.9 per cent is visible at 31.2 d.
The fact that the two methods converge on the same result suggests that it is reliable and that there is a mechanism causing recurrent variations on a roughly monthly period.
The Deeming spectral window \citep{deeming1975} does not show any pathology in the data distribution at the frequency corresponding to this signal.
In Fig.~\ref{acftw} we plot a sinusoidal function with the above period, to be compared with the general trend of the light curve. The agreement is fair, though the occurrence of the observed flares is sometimes slightly shifted with respect to the analysis prediction, indicating that we must rather speak of quasi-periodicity.

\section{Modelling the long-term optical light curve}
\label{sec_mod}
Quasi-periodicities in blazar light curves have been claimed several times \citep[e.g.][]{sillanpaa1988,sillanpaa1996,lainela1999,raiteri2001,ackermann2015,sandrinelli2017,otero2020}
and have been explained in terms of jet precession or orbital motion in a binary system of supermassive black holes, often involving a helical jet \citep[e.g.][]{villata1998,ostorero2004,rieger2004,valtonen2013,sobacchi2017}. 

Many observations suggest that blazar jets can have a bending, helical or twisted structure 
\citep{mchardy1990,conway1995,britzen2005,perucho2012,fromm2013,larionov2013,britzen2017}. 
Magnetohydrodynamical simulations in 3D show that jet wiggling can result from kink instabilities or other mechanisms 
\citep{nakamura2001,moll2008,mignone2010,liska2018}.
A helical jet model was developed by \citet{villata1999} to explain the spectral variability of Mkn~501 and later applied to interpret the multiwavelength behaviour of other blazars \citep{raiteri1999,raiteri2009}.
In particular, in \citet{ostorero2004} the rotation of the helical jet, possibly driven by orbital motion, was able to explain the 5.7~yr quasi-periodicity of AO~0235+164 found by \citet{raiteri2001}.

In agreement with the above studies, we adopt a model where the relativistic emitting plasma continuously flows along the helical structure with a constant bulk Lorentz factor $\Gamma$. The jet is inhomogeneous in the sense that any portion of it, at its fixed distance from the jet apex, emits a constant flux in a given frequency band. This is determined by the local physical properties, like magnetic field, plasma density, optical depth. The higher the synchrotron frequencies emitted, the closer to the jet apex are the corresponding emitting regions. X-rays\footnote{Significant X-ray synchrotron emission is expected only in high-energy peaked BL Lacs, which is not the case of S4~0954+65.}, ultraviolet, optical, infrared, micro and mm waves, radio waves at longer and longer wavelengths are thus produced at increasing distances from the central engine, and along more and more elongated (and consequently more curved) regions downstream the jet. All these jet properties are assumed to remain constant in time, apart from the rotation of the helical structure, which is the only source of observed variability at any frequency. As the helix rotates, every emitting region will change its orientation with respect to the line of sight and this will produce a variation of the observed flux.
Indeed, because of relativistic boosting, the observed flux density $F_\nu$ is enhanced with respect to the rest-frame flux density $F'_{\nu'}$ by some power of the Doppler factor $\delta$. In a continuous jet \citep[e.g.][]{urry1995}: 
\begin{equation}
F_\nu (\nu) = \delta^{2+\alpha} \, F'_{\nu'} (\nu),
\label{dflu}
\end{equation}
where $\alpha$ is the index of the intrinsic, power-law spectrum, and
\begin{equation}
\delta=[\Gamma \, (1-\beta \, \cos \theta)]^{-1}.
\label{df}
\end{equation}
Therefore, the flux density depends on the Doppler factor, which in turn depends on both the viewing angle $\theta$ and the Lorentz factor $\Gamma=(1-\beta^2)^{-1/2}$, where $\beta$ is the plasma bulk velocity in units of the speed of light.
As a consequence, in a twisting jet, the source is observed in a flaring state at a given frequency every time the corresponding jet emitting region becomes closely aligned with the line of sight, causing an increase of the Doppler factor. 
In our model, higher frequencies are coming from jet regions much shorter than the helix pitch and they show larger amplitude variability because they undergo strong and rapid orientation changes. Lower frequencies, e.g.\ radio waves, are emitted from more extended regions, possibly covering several helix pitches, and their flux variations are much smoother and slower.
In this context, we assume that the region emitting the optical radiation of S4~0954+65 is located in a helical jet, whose rotation produces the 31.2 d quasi-periodicity found in the previous section.
This optical emitting region must still be much shorter than the helical pitch, and so has a very small curvature, and consequently shows strong and rapid variability as the helix rotates and its viewing angle changes.

\begin{figure}
	\includegraphics[width=\columnwidth]{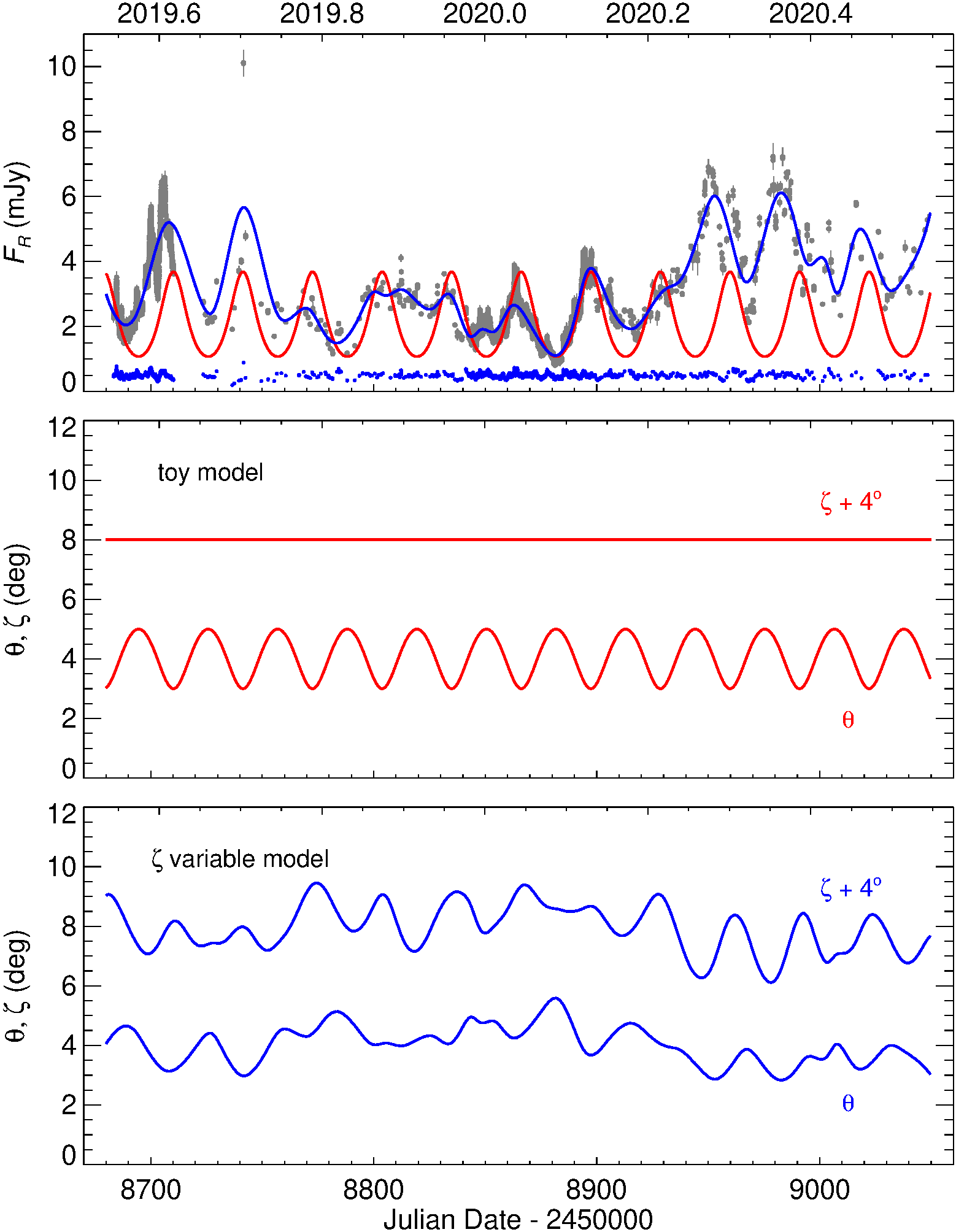}
    \caption{Top: The optical light curve of S4~0954+65 (grey points) with long-term trends overlapped. The red line refers to the toy model with pitch angle $\zeta$ fixed to $4\degr$. The blue line is a cubic spline interpolation through the 7-d binned light curve and also represents the final result of the model with variable $\zeta$. Middle: The periodic behaviour of the viewing angle $\theta$ obtained with the toy model. Bottom: the angles $\theta$ and $\zeta$ resulting from the model with variable $\zeta$ that produce the long-term trend shown in blue in the top panel. Correction of the observed flux densities for the variable Doppler boosting produces the ``residual" fluxes shown in the top panel (blue points). }
    \label{modello}
\end{figure}

In the rotating helical jet scenario, $\theta(t)$ of a given emitting region depends on the helix pitch angle $\zeta$ and on the angle between the helix axis and the line of sight $\psi$ according to
\begin{equation}
\cos\theta(t)=\cos\psi \, \cos\zeta+\sin\psi \, \sin\zeta \, \cos\phi(t).
\label{hel}
\end{equation}
The angle $\phi(t)$ is the variable azimuthal difference between the line of sight and the direction of the bulk plasma velocity in the considered jet region, and contains the information on the periodicity:
$\phi (t)=\omega (t-t_0)=2 \pi (t-t_0)/T$, being $T$ the period and $t_0$ the time of one of the periodic peaks in the observed light curve.
From equation~\ref{hel} we can calculate $\theta(t)$ once $\psi$ and $\zeta$ are set.
We fix $\psi=1 \degr$ and consider a toy model where $\zeta$ is fixed to $4 \degr$.
As shown in Fig.~\ref{modello}, the result is a periodic variation of $\theta(t)$.
From $\theta(t)$, and adopting $\Gamma=10$, we calculate $\delta(t)$ with equation~\ref{df} and then the long-term trend predicted by the toy model from equation~\ref{dflu}.
This is plotted in Fig.~\ref{modello} (with $\alpha=1.82$ as found in Section~\ref{sec_col} from the colour analysis) and shows a general agreement with the observed flux-density behaviour, but cannot reproduce in particular the difference in flare amplitudes.

The fact that the observed flux variations do not appear as perfectly periodic, identical and symmetric outbursts implies that the toy model is an oversimplification and that the helical pattern cannot be a perfect and rigid structure. Indeed, we can expect at least some oscillations around its equilibrium configuration.
The rotation of the jet structure drives the jet helical pattern and tends to create a stronger and stronger toroidal magnetic field component as the helix becomes more twisted. The tension of these magnetic field lines is what collimates the jet and prevents its inflation because of the centrifugal force pushing the helically-moving plasma outwards. This force, added to the various (material and magnetic) pressures acting in the jet, must be in equilibrium with the toroidal magnetic tension\footnote{For an analytical view of helical magnetohydrodynamic equilibria see e.g.\ \citet{villata1993,villata1994a,villata1994b,villata1994c,villata1995}}. Therefore, when the helical jet becomes too twisted (larger pitch angle and stronger magnetic tension), the centrifugal force and the pressures increase too, expanding the structure towards a more relaxed morphology with a smaller pitch angle, and so on around the dynamical equilibrium configuration.

In line with the above scenario, we assume that the pitch angle $\zeta$ can change in time
and calculate numerically its temporal evolution from equation~\ref{hel}.
There $\theta(t)$ is obtained by using equations~\ref{df} and \ref{dflu}, where the flux long-term trend is modelled by means of a cubic-spline interpolation through the 7-d binned light curve (also shown in Fig.~\ref{modello}).
The result is plotted in Fig.~\ref{modello}, where both $\zeta(t)$ and $\theta(t)$ are displayed.
The predicted time behaviour of these angles is what determines the long-term trend traced by the spline interpolation in our geometrical model. 
The corresponding $\delta$ values vary between $\sim 10$ and $\sim 16$.

Once we correct the observed flux densities for the effect of the time-dependent beaming\footnote{This correction is obtained by dividing the flux densities by the long-term trend.}, the light curve of the residual flux densities (shown in Fig.~\ref{modello}) is characterized by smaller fluctuations similar to those obtained from the 5-d detrending in Section~\ref{sec_stv}.

We finally stress that the set of model parameters adopted here are reasonable values that are not necessarily unique, but can demonstrate the strength of the rotating helical jet scenario.
However, it is remarkable that \citet{jorstad2017} found  $\Gamma = 11.4 \pm 3.1$,  $\theta=1.5 \degr \pm 0.7 \degr$ and opening angle $\sim 3.5 \degr$ for the parsec-scale radio jet of S4~0954+65.

\section{Conclusions}
\label{sec_fine}
In this paper we have presented the results of one-year monitoring of S4~0954+65 by the WEBT, complemented by three periods of {\it TESS} observations. The source was in a state of moderate activity, with a maximum variability amplitude of $\sim 2.8$ mag in the $R$ band. In particular, we detected an extreme variability episode, with a brightening of 1 mag in 24 h, followed by a 0.8-mag dimming in 23 h.
The optical colours showed a very mild bluer-when-brighter long-term trend, while the short-term flux changes are much more chromatic, as already found, e.g., in a previous study of another BL Lac object, S5~0716+714, by \citet{raiteri2021}. 
The radio flux looked rather stable at long wavelengths, while it was quite variable at 37 GHz. We found a possible three-weeks time delay of the flux variations at this frequency with respect to the optical ones. 

The source also showed strong variability in both the polarization degree and electric vector polarization angle. There is no general correlation between the $P_{\rm opt}$ and the optical flux.
However, the mean trends of $P_{\rm opt}$ and $\rm EVPA_{opt}$ suggest that the polarization angle changes the direction of rotation every time $P_{\rm opt}$ reaches a maximum or a minimum. 
Although sampled in a sparser way, the radio polarization in general shows a stabler behaviour. 

We conducted a detailed analysis of the short-term flux variations occurring in the {\it TESS} light curves. We applied the detrending method developed by \citet{raiteri2021}, correcting the {\it TESS} light curves for long-term trends with progressively higher time resolution and then calculating the corresponding autocorrelation functions. ACF minima which are common to various detrended light curves are considered to indicate genuine variability time-scales, while common maxima that repeat at nearly equally-spaced time intervals represent possible quasi-periodicities, whose reliability is tested with the periodogram. We found a different behaviour in the three {\it TESS} sectors, with multiple variability time-scales, which are confirmed also by a structure function analysis, but no trustworthy periodicity.

On the contrary, when we examine the long-term composite light curve built with WEBT and {\it TESS} data, we found that the main peaks (and dips) repeat with about one month time interval. This result is confirmed by both the ACF and periodogram analyses. Most notably, in at least two cases we can recognize ``twin" flares (3-4 and 5-6 in Fig.~\ref{acftw}) with the same flux amplitude.

We explain the long-term quasi-periodic variations of the optical light curve of S4~0954+65 in terms of an inhomogeneous helical jet that rotates and whose pitch angle varies in time because of small oscillations of the helical structure around its dynamical equilibrium configuration, which in our case is given by $\zeta \sim 4 \degr$.

In Fig.~\ref{modello} we showed the residual fluxes once the long-term trend due to the variation of the Doppler factor has been removed. 
Their fluctuations are similar to those analysed in the {\it TESS} light curves in Section~\ref{sec_stv}. The variability time-scales of the order of a few days can be explained in terms of a stranded jet structure, as in the S5~0716+714 case studied by  \citet{raiteri2021}. Their geometric nature is consistent with the almost achromatic behaviour. 

Moreover, if the helical structure of the jet is composed of wrapped filaments, then the magnetic field will not follow the main helical path, but will have a more complex structure. This is consistent with the polarimetric data: in the case of a rotating helical structure of the magnetic field, we would expect to see loops of rotation (with different radii in the case of variable pitch angle) of the relative Stokes parameters in a $u$ versus $q$ plot, which are not observed. 

On even shorter time-scales (less than one day), the strongly chromatic variability is likely intrinsic, being associated with energetic processes in the jet, like shock waves in a turbulent plasma. 
Indeed, the superposition of systematic (Doppler factor changes, shocks-in-jet) and stochastic (turbulence) processes can also account for the observed variations of the polarization degree and position angle \citep{weaver2020}.

In this scenario, the smoothness and time delay of the radio emission would indicate that most of it comes from a jet region that is spatially separated (likely downstream) and wider than the optical one. 
In this case, the radio emission would be the result of the sum of the contributions coming from a broad continuous range of viewing angles \citep[e.g.][]{raiteri2017_nature}, so that the marking of the optical periodicity is swamped. 
Finally, according to this model, the radio jet components visible in the VLBI/VLBA images of this and other sources  \citep[e.g.][]{morozova2014,casadio2015,jorstad2017,larionov2020} would be the most beamed parts of the continuous helical path \citep[e.g.][]{bach2006}.

\section*{Acknowledgements}
We dedicate this paper to the memory of our esteemed colleague Valeri M. Larionov, who has been a WEBT pillar for 20 years and whose loss we deeply mourn.
We thank the SRT operators Elise Egron, Delphine Perrodin and Mauro Pili for help with the observations.
This paper includes data collected by the {\it TESS} mission. Funding for the {\it TESS} mission is provided by the NASA Explorer Program.
Partly based on observations made with the Nordic Optical Telescope, owned in collaboration by the University of Turku and Aarhus University, and operated jointly by Aarhus University, the University of Turku and the University of Oslo, representing Denmark, Finland and Norway, the University of Iceland and Stockholm University at the Observatorio del Roque de los Muchachos, La Palma, Spain, of the Instituto de Astrofisica de Canarias.
This article is partly based on observations made in the Observatorios de Canarias del IAC with the Liverpool telescope operated on the island of La Palma by the Liverpool John Moores University in the Observatorio del Roque de Los Muchachos. This article is partly based on observations made with the LCOGT Telescopes, one of whose nodes is located at the Observatorios de Canarias del IAC on the island of Tenerife in the Observatorio del Teide.
This article is partly based on observations made with the IAC-80 operated on the island of Tenerife by the Instituto de Astrofisica de Canarias in the Spanish Observatorio del Teide. Many thanks are due to the IAC support astronomers and telescope operators for supporting the observations at the IAC-80 telescope.
This publication makes use of data obtained at Mets\"ahovi Radio Observatory, operated by Aalto University in Finland.
This research has made use of NASA's Astrophysics Data System and of the NASA/IPAC Extragalactic Database (NED), which is operated by the Jet Propulsion Laboratory, California Institute of Technology,
under contract with the National Aeronautics and Space Administration.
The research at Boston University was supported by NASA grants 80NSSC20K1566(Fermi Guest Investigator Program) and 80NSSC21K0243 (TESS Guest Investigator Program). This study was based (in part) on observations conducted using the 1.8 m Perkins Telescope Observatory (PTO) in Arizona (USA), which is owned and operated by Boston University.
G.D., M.S., G.M. and O.V. acknowledge the observing grant support from the Institute of Astronomy and Rozhen NAO BAS through the bilateral joint research project "Gaia Celestial Reference Frame (CRF) and fast variable astronomical objects" (2020-2022, leader is G.Damljanovic), and support by the Ministry of Education, Science and Technological Development of the Republic of Serbia (contract No 451-03-68/2020-14/200002)
This research was partially supported by the Bulgarian National Science Fund of the Ministry of Education and Science under grants DN 18-13/2017, KP-06-H28/3 (2018), KP-06-H38/4 (2019) and KP-06-KITAJ/2 (2020).
S.O.K. acknowledges financial support by Shota Rustaveli National Science Foundation of Georgia under contract PHDF-18-354
E.B. acknowledges support from DGAPA-PAPIIT GRANT IN113320.
This work is partly based upon observations carried out at the Observatorio Astron\'omico Nacional on the Sierra San Pedro M\'artir (OAN- SPM), Baja California, Mexico.
We acknowledge support by Bulgarian National Science Fund under grant DN18-10/2017 and National RI Roadmap Project DO1-383/18.12.2020 of the
Ministry of Education and Science of the Republic of Bulgaria.
I.A. acknowledges financial support from the Spanish ``Ministerio de Ciencia e Innovaci\'on" (MCINN) through the ``Center of Excellence Severo Ochoa" award for the Instituto de Astrof\'isica de Andaluc\'ia-CSIC (SEV-2017-0709). Acquisition and reduction of the POLAMI data was supported in part by MICINN through grants AYA2016-80889-P and PID2019-107847RB-C44. The POLAMI observations were carried out at the IRAM 30m Telescope. IRAM is supported by INSU/CNRS (France), MPG (Germany) and IGN (Spain).

\section*{Data availability}
The data collected by the WEBT collaboration are stored in the WEBT archive at the Osservatorio Astrofisico di Torino - INAF (http://www.oato.inaf.it/blazars/webt/); for questions regarding their availability, please contact the WEBT President Massimo Villata ({\tt massimo.villata@inaf.it}).




\bibliographystyle{mnras}
\bibliography{pawete} 


\section*{Affiliations}
{\it
$^{ 1}$INAF, Osservatorio Astrofisico di Torino, via Osservatorio 20, I-10025 Pino Torinese, Italy                                                           \\
$^{ 2}$Astronomical Institute, St. Petersburg State University, Universitetskij Pr. 28, Petrodvorets, St. Petersburg 198504, Russia                          \\
$^{ 3}$Pulkovo Observatory, St.-Petersburg, 196140, Russia                                                                                                   \\
$^{ 4}$Institute for Astrophysical Research, Boston University, 725 Commonwealth Avenue, Boston, MA 02215, USA                                               \\
$^{ 5}$Instituto de Astrof\'{i}sica de Canarias, E-38200, La Laguna, Tenerife, Spain                                                                         \\
$^{ 6}$Universidad de La Laguna, Departamento de Astrofisica, E38206, La Laguna, Tenerife, Spain                                                             \\
$^{ 7}$Instituto de Astrof\'isica de Andaluc\'ia-CSIC, Glorieta de la Astronom\'ia s/n, E—18008, Granada, Spain                                            \\
$^{ 8}$Institute of Applied Astronomy RAS, Nab Kutuzova 10, 191187 St Petersburg, Russia                                                                     \\
$^{ 9}$Institute of Astronomy and National Astronomical Observatory, Bulgarian Academy of Sciences, 72 Tsarigradsko Shose Blvd., Sofia, Bulgaria             \\
$^{10}$Universidad Nacional Aut\'onoma de M\'exico, Instituto de Astronom\'ia, AP 70-264, CDMX  04510, Mexico                                                \\
$^{11}$Finnish Centre for Astronomy with ESO (FINCA), University of Turku, Quantum, Vesilinnantie 5, 20014 University of Turku, Finland                      \\
$^{12}$Aalto University Mets\"ahovi Radio Observatory, Mets\"ahovintie 114, 02540 Kylm\"al\"a, Finland                                                       \\
$^{13}$Aalto University Department of Electronics and Nanoengineering, P.O. BOX 15500, FI-00076 AALTO, Finland                                               \\
$^{14}$Crimean Astrophysical Observatory RAS, P/O Nauchny, 298409, Crimea                                                                                    \\
$^{15}$Department of Astronomy, Faculty of Physics, University of Sofia, BG-1164, Sofia, Bulgaria                                                            \\
$^{16}$EPT Observatories, Tijarafe, E-38780 La Palma, Spain                                                                                                  \\
$^{17}$INAF, TNG Fundaci\'on Galileo Galilei, E-38712 La Palma, Spain                                                                                        \\
$^{18}$Foundation for Research and Technology - Hellas, IESL \& Institute of Astrophysics, Voutes, 7110 Heraklion, Greece                                    \\
$^{19}$Department of Physics, University of Crete, 71003, Heraklion, Greece                                                                                  \\
$^{20}$Max Planck Institute for Radio Astronomy, Auf dem H\"ugel, 69 D-53121 Bonn, Germany                                                                  \\
$^{21}$Graduate Institute of Astronomy, National Central University, Zhongli 32051, Taoyuan, Taiwan                                                          \\
$^{22}$Astronomical Observatory, Volgina 7, 11060 Belgrade, Serbia                                                                                           \\
$^{23}$INAF, Istituto di Radioastronomia, Via Gobetti 101, I-40129 Bologna, Italy                                                                            \\
$^{24}$Aryabhatta Research Institute of Observational Sciences (ARIES), Manora Peak, Nainital - 263001, India                                                \\
$^{25}$Universidad Nacional Aut\'onoma de M\'exico, Instituto de Astronom\'ia, AP 106, Ensenada 22860, BC, Mexico                                            \\
$^{26}$Korea Astronomy and Space Science Institute, 776 Daedeok-daero, Yuseong-gu, Daejeon 34055, Republic of Korea                                          \\
$^{27}$Abastumani Observatory, Mt. Kanobili, 0301 Abastumani, Georgia                                                                                        \\
$^{28}$Nordic Optical Telescope, Apartado 474, E-38700 Santa Cruz de La Palma, Spain                                                                         \\
$^{29}$Observatoire de Gen\`eve, Universit\'e de Gen\`eve, Chemin Pegasi 51, 1290, Versoix, Switzerland                                                      \\
$^{30}$Landessternwarte, Zentrum f\"ur Astronomie der Universit\"at Heidelberg, K\"onigstuhl 12, D-69117 Heidelberg, Germany                                 \\
$^{31}$Engelhardt Astronomical Observatory, Kazan Federal University, Tatarstan, Russia                                                                      \\
$^{32}$Samtskhe-Javakheti State University, 92 Shota Rustaveli St. Akhaltsikhe, Georgia                                                                       \\
$^{33}$Faculty of Mathematics, University of Belgrade, Studentski trg 16, 11000 Belgrade, Serbia                                                             \\
$^{34}$Instituto de Radioastronom\'ia Milim\'etrica, Avenida Divina Pastora, 7, N\'ucleo Central, E—18012, Granada, Spain                                  \\
$^{35}$Department of Physics and Mathematics, Aoyama Gakuin University, 5-10-1 Fuchinobe, Chuo-ku, Sagamihara, Kanagawa 252-5258, Japan                      \\
$^{36}$Special Astrophysical Observatory, Russian Academy of Sciences, 369167, Nizhnii Arkhyz, Russia                                                        \\
}


\appendix

\section{Stucture Function analysis}
\begin{figure}
	\includegraphics[width=\columnwidth]{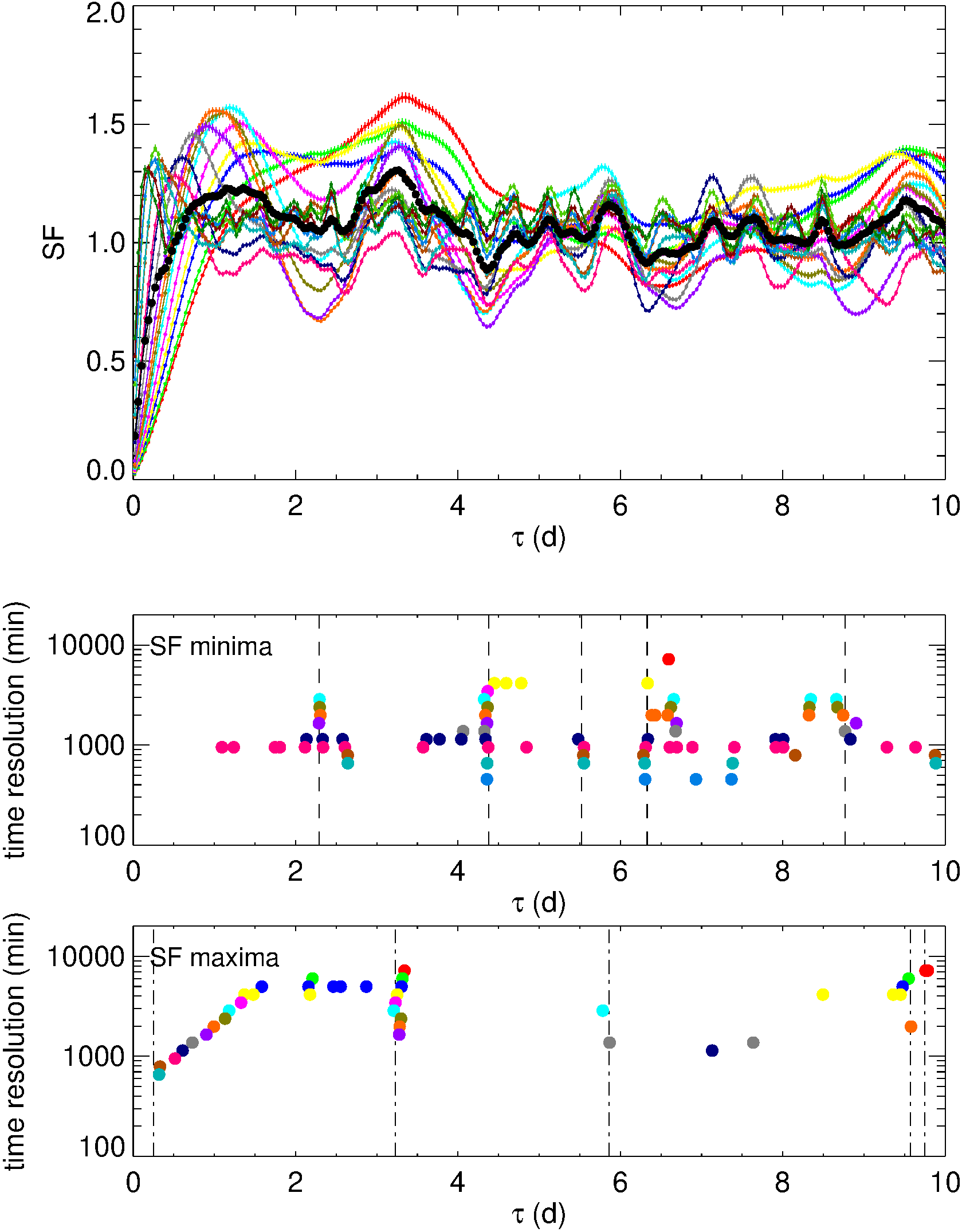}
    \caption{Top panel: the SFs (normalized to their mean value) of the detrended light curves of Sector 14 shown in Fig.~\ref{detre14}. The black dots represent the average SF, whose maximum and second minimum are used as thresholds for the significance of the maxima/minima of the single SFs. Middle/Bottom panel: the time lag of the most significant SF minima/maxima versus the time resolution of the spline used to perform the detrending. Vertical lines mark the results obtained with the ACFs in Section~\ref{sec_acf}.}
    \label{sf14}
\end{figure}

\begin{figure}
	\includegraphics[width=\columnwidth]{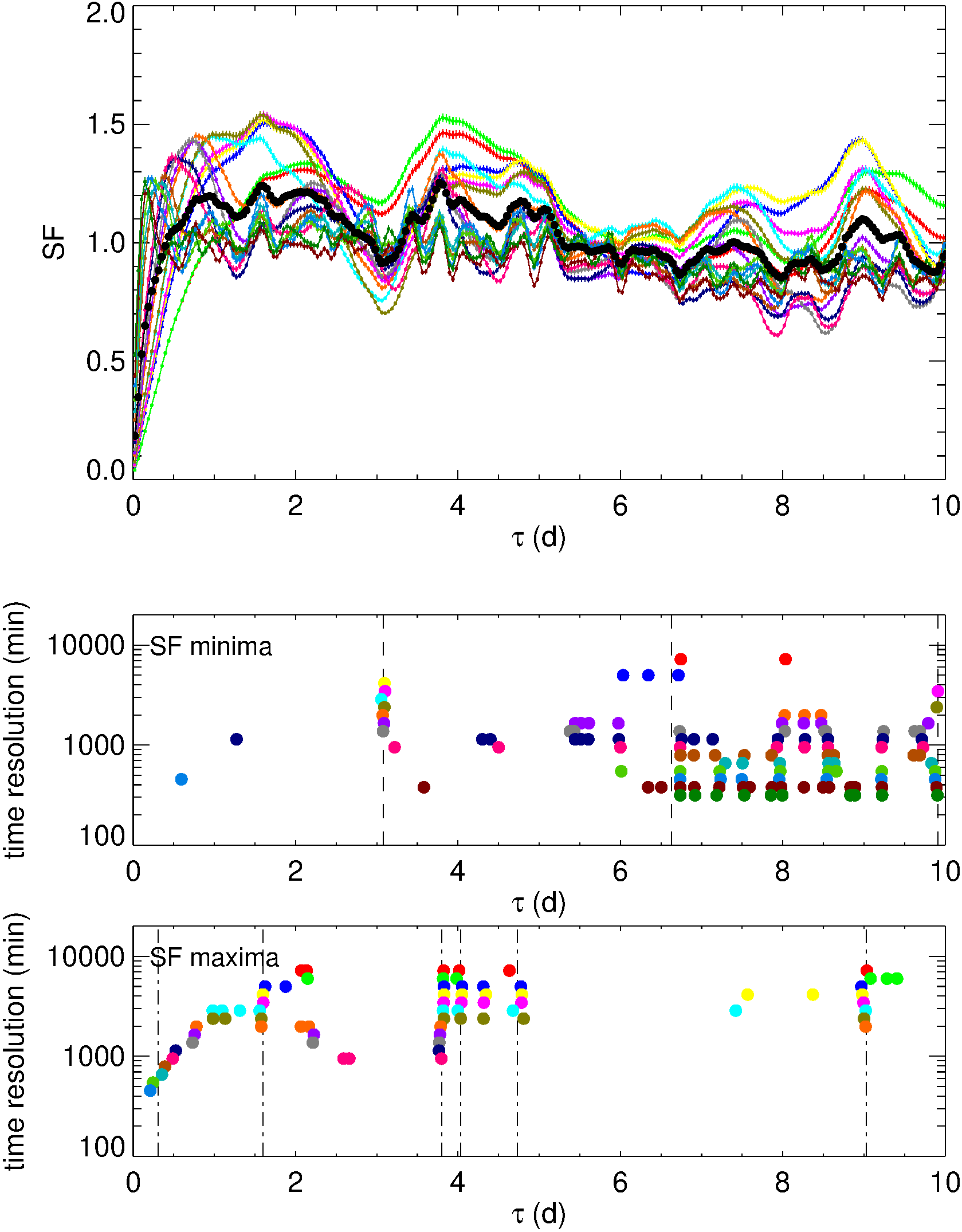}
    \caption{As in Fig.~\ref{acf14}, but for Sector 20.}
    \label{sf20}
\end{figure}

\begin{figure}
	\includegraphics[width=\columnwidth]{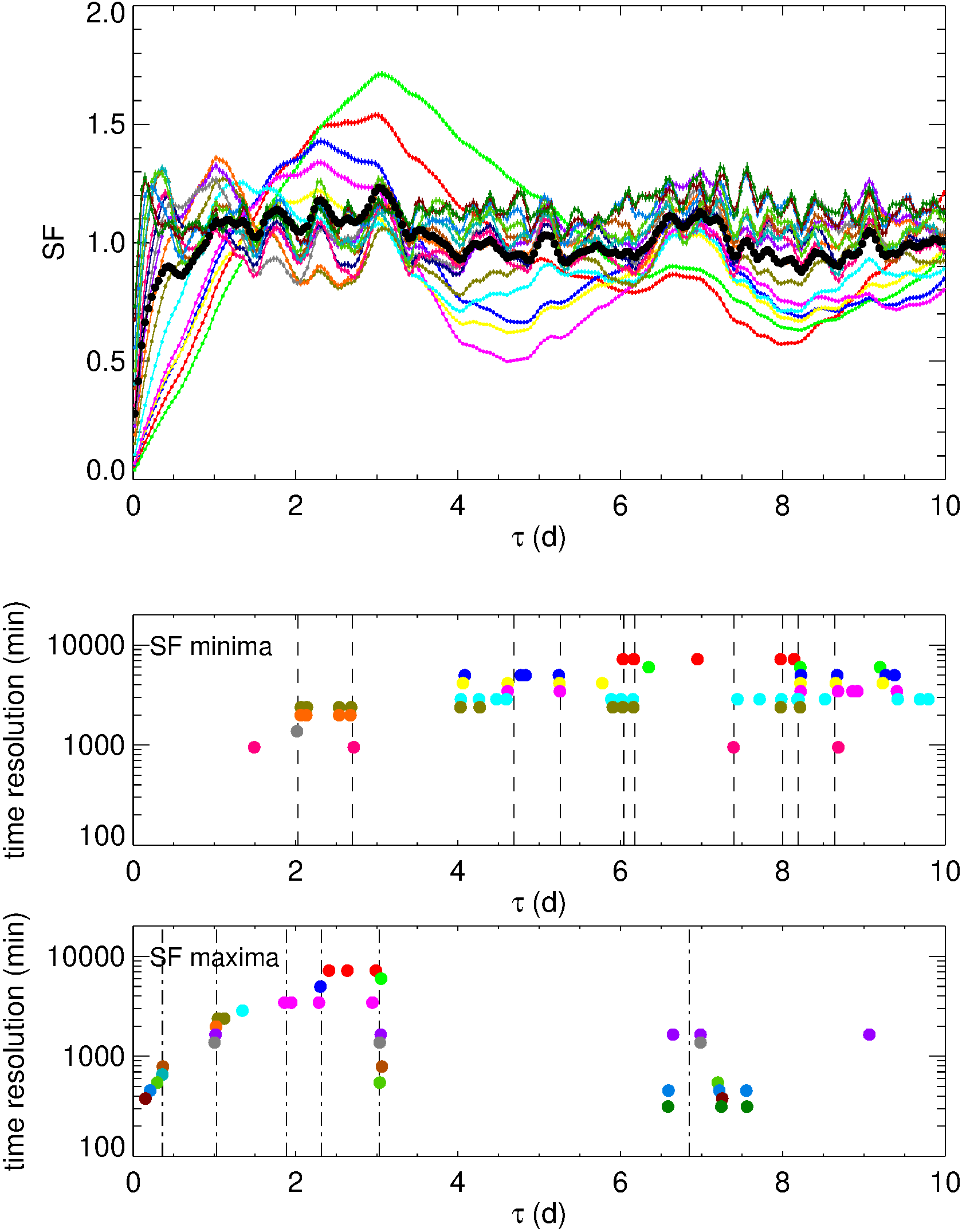}
    \caption{As in Fig.~\ref{acf14}, but for Sector 21.}
    \label{sf21}
\end{figure}

The SF is here used to validate the results obtained by the ACF in Section~\ref{sec_acf}.
Figs.~\ref{sf14}, \ref{sf20} and \ref{sf21} show the SFs obtained for the detrended light curves shown in Figs.~\ref{detre14}, \ref{detre20} and \ref{detre21}, respectively.
As in the ACF case, we also calculated the mean SF and used its maximum and second minimum to set thresholds to identify the most significant SF maxima and minima in the single SFs. These minima/maxima thresholds are 0.945/1.260 for Sector~14, 0.935/1.191 for Sector~20, and 0.961/1.217 for Sector~21. 
The comparison of the SF minima with the ACF maxima and of the SF maxima with the ACF minima reveal a good agreement, because all ACF time-scales are confirmed by the SFs. 
Moreover, the SFs reveal some additional signals: some of them were already present in the ACFs, but did not fulfill the requirement to be present in more than three detrended light curves.

\bsp	
\label{lastpage}
\end{document}